\documentclass[11pt,a4paper]{article}
\pdfoutput=1
\usepackage[latin1]{inputenc}
\usepackage[T1]{fontenc}
\usepackage[english]{babel}
\usepackage{bm}
\usepackage{bbold}
\usepackage{textcomp}
\usepackage{url}
\usepackage{slashed}
\usepackage{amsmath, mathbbol}
\usepackage{graphicx}
\usepackage{color}
\usepackage{tabularx}
\usepackage{amssymb}
\usepackage{bbding}
\usepackage{multirow}
\usepackage{subfigure}

\newcommand{\non}[0]{\nonumber \\}
\newcommand{\bee}[0]{\begin{eqnarray}}
\newcommand{\eee}[0]{\end{eqnarray}}

\newcommand{\be}[0]{\begin{equation}}
\newcommand{\ee}[0]{\end{equation}}

\renewcommand{\tilde}{\widetilde}

\usepackage{multirow}
\makeatletter

\def\lsim{\;\raise0.3ex\hbox{$<$\kern-0.75em\raise-1.1ex\hbox{$\sim$}}\;}
\def\gsim{\;\raise0.3ex\hbox{$>$\kern-0.75em\raise-1.1ex\hbox{$\sim$}}\;}

\def\ben{\begin{enumerate}}  \def\een{\end{enumerate}}
\def\bit{\begin{itemize}}    \def\eit{\end{itemize}}
\def\beq{\begin{equation}}   \def\eeq{\end{equation}}
\def\ba{\begin{array}}       \def\ea{\end{array}}
\def\bea{\begin{eqnarray}}   \def\eea{\end{eqnarray}}

\begin{document}
\begin{titlepage}
\vspace*{-1cm}
\flushright{LPT-Orsay 13-39\\SISSA  59/2013/FISI}
\vspace*{1.5cm}
\begin{center}
{\Large\bf Looking for the minimal inverse seesaw realisation
}
\end{center}
\vskip 0.7  cm
\begin{center}
{\large A. Abada$\,^{a}$\footnote{asmaa.abada@th.u-psud.fr} and M. Lucente}$\,^{a,b}$\footnote{michele.lucente@th.u-psud.fr} \\\vskip 0.4  cm
$^a\,$ Laboratoire de Physique Th\'eorique,\\
Universit\'e de Paris-Sud 11, B\^at. 210, 91405 Orsay Cedex, France\\

$^b\,$ Scuola Internazionale Superiore di Studi Avanzati,\\
via Bonomea 265, 34146 Trieste, Italy

\end{center}
\vskip 0.5cm
\begin{abstract}
\noindent  In this work we consider a simple extension of the Standard Model  involving additional fermionic singlets and assume an underlying inverse seesaw mechanism (with one or more right-handed neutrinos and one or more  sterile fermions) for neutrino mass generation.  Under the assumption that both sterile states and right-handed neutrinos are present, our goal is to  determine which is  the minimal inverse seesaw realisation that accounts for neutrino data while at the same time complying with all experimental  requirements (electroweak precision tests and laboratory constraints).
This study aims at identifying  the minimal inverse seesaw realisation for the 3-flavour and for the 3 +~more-mixing schemes, the latter giving an explanation for the reactor anomalies and/or providing a possible candidate for the dark matter of the Universe.  Based on a perturbative approach, our generic study shows that in the class of inverse seesaw models giving rise to a 3-flavour flavour mixing scheme, only two mass scales are relevant (the light neutrino mass scale, $m_\nu$ and the mass of the right-handed neutrinos, $M_R$) while in the case of a 3 + 1-mixing scheme, an additional mass scale ($\mu$ $\in[m_\nu,M_R]$) is required.
For each of the two obtained  inverse seesaw frameworks,  we conduct a thorough numerical analysis,  providing predictions for the hierarchy of the light neutrino spectrum and for the effective mass in neutrinoless double beta decay. 

\end{abstract}
\end{titlepage}
\setcounter{footnote}{0}
\vskip2truecm
\newpage
\setcounter{page}{2}
\section{Introduction}\label{Sec:intro}
Oscillation experiments have established a clear  evidence for two oscillation frequencies ($\Delta m_{ij}^2$) - implying that at least two neutrino states are massive - as well as  the basic structure of a 3-flavour leptonic mixing matrix  (for a recent review, see~\cite{GonzalezGarcia:2012sz}). However,  
current  reactor~\cite{reactor:I}, accelerator~\cite{Aguilar:2001ty,miniboone:I}  and Gallium anomalies~\cite{gallium:I} suggest that there might be some extra fermionic gauge singlets with mass(es) in the eV range.
This would  imply  that instead of  the three-neutrino mixing scheme, one would have a $3\ +\ 1$-neutrino (or $3\ +$ more) mixing schemes (for a global overview, see~\cite{Kopp:2013vaa}).

Sterile fermionic states (not necessarily light) are present in several neutrino mass models and their masses can range from well below the electroweak scale up to the Planck scale.  Other than the reactor and accelerator anomalies, the existence of sterile states is also motivated  by certain indications from large scale structure formation~\cite{Kusenko:2009up,Abazajian:2012ys}. 
Nevertheless, due to the mixings of the sterile fermionic states with the active left-handed neutrinos,  models with sterile fermions are severely constrained from electroweak (EW) precision observables, laboratory data and cosmology. 

In contrast with the huge experimental achievements in determining neutrino oscillation parameters,  
 many questions remain to be answered concerning neutrino properties, as  for instance  the neutrino nature (Majorana or Dirac),  the  absolute neutrino mass scale and  the hierarchy of the neutrino mass spectrum, which are not yet determined. Finally, and most importantly, one must unveil the neutrino mass generation mechanism at work and which  new physics scales are required.

One of the most economical possibilities to account for massive neutrinos is to embed a standard seesaw 
mechanism (of type I, II or III)~\cite{seesaw:I, seesaw:II, seesaw:III} 
into the framework of the Standard Model (SM). The caveat of these scenarios is that, in
order to have natural neutrino Yukawa couplings  
the typical scale of the extra particles
(such as right-handed neutrinos, scalar or fermionic isospin triplets) is in general very high, 
potentially very close to the gauge coupling unification (GUT) scale, thus implying that direct experimental tests of
the seesaw hypothesis might be  impossible.  
In contrast, low-scale seesaw mechanisms~\cite{Mohapatra:1986bd,GonzalezGarcia:1988rw,Deppisch:2004fa,Asaka:2005an,Gavela:2009cd,Ibarra:2010xw}, in which sterile fermions are added to the SM particle content with masses
around the electroweak scale or even lower, are very attractive from a
phenomenological point of view since the new states can be
produced in collider and/or low-energy experiments, and their
contributions to physical processes can be sizeable.

In view of the strong potential of low-scale seesaw mechanisms, in this work we consider the inverse seesaw (ISS) 
mechanism~\cite{Mohapatra:1986bd,GonzalezGarcia:1988rw,Deppisch:2004fa} which requires the addition of both  $\#\nu_R\ne 0$ right-handed (RH) neutrinos 
and $\# s\ne 0$ extra sterile fermions to the SM field content\footnote{In the case where $\# s=0$, one recovers the type I  seesaw realisation which could account for neutrino masses and mixings provided that the number of right-handed neutrinos is at least $\#\nu_R=2$.}.
The distinctive feature of the ISS is that an additional dimensionfull  
parameter ($\mu$) allows to accommodate the smallness of the active neutrino
masses $m_\nu$ for a low seesaw scale, and still with natural Yukawa
couplings ($Y^\nu\sim {\mathcal{O}}(1)$).  In turn, this allows for 
sizeable mixings between the active and the additional
sterile states. Such features are in clear contrast with, for instance, the canonical type I seesaw~\cite{seesaw:I},
where $\mathcal{O}(1)$ Yukawa couplings require the mass of the right-handed neutrinos to be close to the GUT scale, 
$M_R \sim 10^{15}$~GeV, thus leading to extremely small active-sterile mixings.

Any type I seesaw realisation requires the introduction of $N$ gauge singlet Weyl fermions $w$ that can thus couple via a Majorana mass term $\sim M_{ij} w^c_i w_j$. Both the number $N$ and the energy scale $M$ are in principle free parameters that can be fixed by neutrino data. It is thus natural to ask what is the minimal number of fermionic singlets  $N$ required to successfully generate neutrino masses and mixings in agreement with experiments.  It was shown in~\cite{Donini:2011jh} that the choice $N=1$, although containing in principle enough parameters, fails in fitting all neutrino oscillation experiments, while the choice $N=2$ is the minimal one that is phenomenologically viable. If no structure is assumed for the $M$ matrix,  type I seesaw realisations usually contain only one relevant energy scale related to the mass of the new sterile fermions. If they can be integrated out, this energy scale $M$, which suppresses the dimension 5 effective operator (actually responsible for the smallness of neutrino masses), double-suppresses the dimension 6 operator  that can induce lepton flavour violating (LFV) processes. In  this situation tiny neutrino masses necessarily imply very strongly suppressed LFV processes. As pointed out in~\cite{Abada:2007ux}, the situation is different for instance in the type II seesaw, or in the type I seesaw, when the matrix $M$ has some specific structure leading  for example to the inverse seesaw scenario.
In this case,  the matrix $M$ exhibits two different energy scales (as a consequence of the lepton number assignment  of the new singlet Weyl fermions), among them explicit total lepton number violating (LNV) entries, very small compared to the conserving ones.
This implies that the same high-energy suppression is expected for the dimension 5 and 6 effective operators, but the former one is further suppressed by the small LNV parameters. It is thus possible to generate tiny neutrino masses and sizeable coefficients for the dimension 6 LFV operators. Minimal models in this framework have been addressed in~\cite{Gavela:2009cd}, where $N=2n$ Weyl fermions were added to the SM field content with a  lepton number assignment allowing them to be cast into two groups of $n$ elements with opposite lepton number charges. It was found that the minimal phenomenologically viable model is the one with $n=1$, which can be the mechanism at work  if all the (gauge invariant) lepton number violating interactions are allowed.  In this situation the tree level neutrino masses derive from the sum of two terms which are differently suppressed by the high-energy scale - and which  depend on  two sets of Yukawa couplings present (lepton number violating and conserving) - while the coefficients of the LFV dimension 6 operators only  depend,  to a first approximation, on the lepton number conserving Yukawas. 
The situation is different in the case of the inverse seesaw scenario, where LNV Yukawas are not allowed and the dimension 5 and 6 effective operators have the same high-energy suppression~\cite{Gavela:2009cd}. The price to pay in this case is that the minimal phenomenologically viable model is the one with $n=2$, that is $N=4$. 

Usually, in the  inverse seesaw scenario, where a LNV parameter $\mu$ is present,  an equal number of singlet Weyl fermions with opposite lepton number is added to the SM field content, i.e. $N=n+n$.  After the diagonalisation of the neutral mass matrix, one ends up with three active neutrinos (at least  two massive in order to accommodate neutrino data) and  $n$ pseudo-Dirac pairs with mass differences of the order of the LNV parameter $\mu$. Notice that in this scenario  the scale $\mu$ does not correspond to the mass of any new physical state (after diagonalisation). 
In this work,  we will consider  the inverse seesaw scenario in which  we relax  the previous assumption: adding $N=n+n'$ Weyl fermions with opposite lepton number,  with $n$  not necessarily coinciding with $n'$. 
We will show that when $n\ne n'$, the LNV scale $\mu$ can indeed correspond to the mass of a physical (almost sterile) state, i.e.,  a light sterile neutrino.

Since both RH neutrinos and sterile states are gauge singlets, there is no requirement on their (generation) number from anomaly cancellation.
Moreover, in view of the presence of two independent mass scales 
(the mass of the RH neutrinos and the Majorana mass of the sterile states), associated to gauge singlet fermions, it is only natural to investigate which is the 
minimal content of the ISS extension of the SM successfully accounting for neutrino data, while at the same time complying with all available experimental and observational constraints. 

We thus embed the inverse seesaw  mechanism into the SM, considering
models with an arbitrary non-vanishing (and different) number of RH neutrinos and of additional sterile states, in order to establish which class of models provides a  minimal 3-flavour and   3 +~more-mixing schemes. The latter class of realisations (configurations) may 
offer an explanation to the  reactor anomalies or, depending on the mass scales, a (partial) solution for the Dark Matter (DM) problem, in the form of a Warm DM (WDM) candidate~\cite{deVega:2013ysa}. 
In a first stage, we do not impose a particular mass  scale for the (RH) Majorana states nor the hierarchy of the associated light spectrum; 
likewise, we do not specify a mass range for the sterile fields. 

Our study has allowed to identify two
classes of minimal ISS realisations that can successfully account for neutrino data:  
the first leads to a 3-flavour mixing scheme, and requires only two scales (that of light neutrino masses, $m_\nu$, and the mass of the RH neutrinos, $M_R$); 
the second corresponds to a 3~+~1-mixing scheme, and calls for an additional scale 
($\mu$ $\in[m_\nu,M_R]$).  
For each of these minimal classes, we carried  a numerical analysis taking into account all possible bounds associated to the presence of sterile fermions (which 
constrain the mixings between active and sterile neutrinos for different mass regimes). 
We  also provide predictions regarding the hierarchy of the light neutrino spectrum (normal or inverted) and the effective mass in neutrinoless double beta decay, for each of the minimal realisations identified. 

The paper is organised as follows:  in Section~\ref{Sec:towards}, we 
briefly review the inverse seesaw mechanism and define the framework; we also determine the generic class of frameworks leading to 3- and to 3 +~more-mixing schemes as well as their generic features concerning the different mass scales. 
In Section~\ref{Sec:constrains}, we  consider all the different constraints 
from neutrino data, electroweak observables and laboratory measurements 
 applied  in the analysis. Section~\ref{Sec:analysis} 
is devoted
to the phenomenological analysis of the minimal ISS framework leading to the 3-flavour and to the 3~+~1-mixing schemes. 
Our final remarks are given in
Section~\ref{conclusions}.  For completeness, some  technical details concerning the computation
are included in the Appendices.

\section{Towards the minimal inverse seesaw realisation}\label{Sec:towards}
In this work we consider the inverse Seesaw mechanism~\cite{Mohapatra:1986bd,GonzalezGarcia:1988rw,Deppisch:2004fa} for the generation of  neutrino masses and lepton mixings,  with a minimal field content. 
We work in the framework of the SM extended by one or more generations of right-handed neutrinos $\nu_R$ and additional fermionic singlets~$s$. 

\subsection{The one generation case} \label{largeM}
We first consider the illustrative one generation case. 
In the basis $n_L \equiv \left( \nu_L,\nu_R^c,s \right)^T$,  the  neutrino mass term  reads:
\be
-\mathcal{L}_{m_\nu} =\frac{1}{2} n_L^T\ C\ {M}\ n_L + \text{h.c.},
\ee
where $C\equiv i \gamma^2 \gamma^0$ is the charge conjugation matrix and  the matrix ${M}$ is given by
\be\label{isszeromatrix}
{M} = \left( \begin{array}{ccc} 0 & d & 0 \\ d & m & n \\ 0 & n & \mu \end{array} \right).
\ee
We assume that there is no term mixing the left-handed neutrino with  the fermionic singlet $s$ ($ \sim \overline{\nu_L^c} s$). 
In the above, $d$ corresponds to the Dirac mass term. The matrix $ {M}$ also includes  a Majorana mass term  for the RH neutrino,
\be\label{majmassright}
- \frac{m^*}{{2}} \nu_R^T C \nu_R + \text{h.c.}\,.
\ee
The values of $m$ and $\mu$ in Eq.~(\ref{majmassright})  are arbitrary. However, accommodating neutrino masses of $\mathcal{O}(\text{eV})$ implies that both  must be very small in the case of the inverse seesaw framework.
Assigning  a leptonic charge to both $\nu_R$ and $s$, with lepton number
$L=+1$~\cite{Mohapatra:1986bd,GonzalezGarcia:1988rw,Deppisch:2004fa}
 (such that the Dirac mass term $-d^* \overline{\nu_L} \nu_R + \text{h.c.}$ preserves the leptonic number),  the terms  
$\nu_R^T C \nu_R$ and $s^T C s$ 
violate total leptonic number $L$ by two units. 
Small values of $m$ and 
 $\mu$  are  natural in the sense of 't~Hooft~\cite{'tHooft:1979bh} since in the limit where 
$m, \mu\to 0$, the total lepton number symmetry is restored. 
 In the following, we  assume for simplicity  that $\mu$ and $m$ are of the same order of magnitude.

In order to obtain the {tree-level} neutrino mass spectrum and the leptonic mixing, we diagonalize the matrix ${M}$ as \cite{Schechter:1980gr}
 \be\label{diagonalization}
U^T M U = \mbox{diag}(m_0,m_1,m_2)\,,
\ee
where $U$ is a unitary matrix, and $m_{0,1,2}$  
correspond to the physical neutrino masses. The mixing matrix is obtained from
\be\label{diagmsquare}
\mbox{diag}(m_0^2,m_1^2,m_2^2) = \left( U^T M U \right)^\dagger \left( U^T M U \right)  = U^\dagger M^\dagger M U\, ,
\ee
so that the matrix $U$ diagonalizing $M^\dagger M$ is the same as  the one in Eq.~(\ref{diagonalization}).

We determine the neutrino spectrum {\it perturbatively}: 
the perturbations correspond to taking into account the tiny effects of the lepton number violating 
diagonal entries, 
\be\label{DeltaM}
\Delta M = \mbox{diag}(0,m,\mu)\,.
\ee 
The lightest neutrino mass arises from perturbative corrections\footnote{We denote by $(n)$ superscript perturbative corrections of order $n$.} to the zeroth order $m_0=0$ eigenvalue;
the two other states are pseudo-Dirac heavy neutrinos, massive and degenerate.

Concerning $m_0$,
the second order corrections ${m_0^2}^{(2)}$ (the first order one gives vanishing contributions) 
 are given by
\begin{equation}\label{1genneutrinomass(0)}
{m_0^2}^{(2)} \,=\, \frac{|d|^4 |\mu |^2}{\left(|d|^2+|n|^2\right)^2} \, ,
\end{equation}
which reduces to the usual inverse seesaw expression once one assumes $|d| \ll |n|$. 
The first order corrections to ${m_{1,2}^2}^{(0)}= |d|^2+|n|^2$ lift the degeneracy: 
\be\label{1genneutrinomass(1,2)}
\begin{array}{cc}
 {m_1^2}^{(1)} = -\frac{\left|\mu ^* n^2+m |d|^2+m |n|^2\right|}{\sqrt{|d|^2+|n|^2}}\,, 
 & {m_2^2}^{(1)}= \frac{\left|\mu ^* n^2+m |d|^2+m |n|^2\right|}{\sqrt{|d|^2+|n|^2}}\,.
\end{array}
\ee
The corresponding eigenvectors allowing to build the leptonic mixing matrix can be found in  Appendix~\ref{AppendixA}.
Notice that in this approach,  the only assumption on the magnitude of the physical parameters, i.e.  
\bea\label{condition}
|m|, \,|\mu| \ll |d|,\,|n|\ ,\quad (n\ne 0)
\eea
 is driven (and justified) 
by the naturalness criterium. Notice that when $n\to 0$, one recover the simple realisation of the usual type I seesaw, which is not the scenario we consider in this study. 

\subsection{Minimal Inverse Seesaw realisations}\label{miss}
In this section, we build the minimal ISS framework complying with experimental observations. The latter lead to the following requirements: 
\begin{itemize}
\item there are $3$ generations of neutrino fields with $SU(2)_L \otimes U(1)_Y$ gauge interactions 
($\# \nu_L=3$);
\item  there are at least $3$ non-degenerate light mass eigenstates. 
\end{itemize}

We extend the one generation matrix  of Eq.~(\ref{isszeromatrix})
to the case of several 
generations of $\nu_R$ and $s$ fields,  so that $M$ now  reads
\be\label{generalmatrix}
M= \left( \begin{array}{ccc} 0 & d & 0 \\ d^T & m & n \\ 0 & n^T & \mu \end{array} \right)\,,
\ee
 $d,m,n,\mu$ now being complex matrices. 
 Since $M$ is symmetric (due to the Majorana character of the fields), it follows that $m$ and $\mu$ are also symmetric matrices.
 
 A possible choice in Eq. (\ref{generalmatrix}) is to set the matrix $n=0$, such that the singlets $s$ decouple. In this case, the model reduces in practice to the type I seesaw model, already compatible with low-energy data. We will conduct our analysis always assuming the (perturbativity) condition  Eq. (\ref{condition}) and thus considering the matrix $n\ne0$ and its entries always such that $|m|, \,|\mu| \ll |d|,\,|n|$.

In the following, we denote by $\# \nu_L, \#\nu_R$  and $\# s$  (with $\# \nu_R\ne 0$ and $\# s\ne 0$) the number of generations of left-handed, right-handed  and sterile fields, respectively. 
The Dirac mass matrix $d$ arises from the  Yukawa couplings to the Higgs boson $(\tilde{\Phi} =i  \sigma^2 \Phi)$,
\be\label{yukawa}
Y_{\alpha \beta} \, \overline{l_L}^\alpha \,\tilde{\Phi }\, \nu_R^\beta + \text{h.c.}\,,
\ee
where $Y$ is a complex matrix, $l_L^\alpha$ denotes the left-handed (LH)  leptonic doublet, 
\be
l_L^\alpha \,=\, \left( \begin{array}{c} \nu_L^\alpha \\ e_L^\alpha \end{array} \right)\,,
\ee
$\alpha$ and $\beta$ being generation indices. After electroweak symmetry breaking (EWSB), 
the matrix $d$ is given by 
\be\label{diracterms}
d_{\alpha \beta} = \frac{v}{\sqrt{2}} \,Y^*_{\alpha \beta}\,, 
\ee
and its dimension  is 
\be
\mbox{ dim } d = \left( \# \nu_L \times \# \nu_R \right).
\ee
The matrix $n$ describes the lepton number conserving interactions involving  
$\nu^c_R$ and $s$ fields, and its dimension is
\be
\mbox{ dim } n = \left( \# \nu_R \times \# s \right).
\ee
Finally, the dimension of the (symmetric) Majorana mass matrices $m$ and $\mu$ are  given by
\be
\mbox{ dim } m = \left( \# \nu_R \times \# \nu_R \right)\,, \quad
\mbox{ dim } \mu = \left( \# s \times \# s \right)\,.
\ee

Being gauge singlets, and since there is no direct evidence for their existence,  the number of additional fermionic 
singlets $\# \nu_R$ and $\# {s}$ is unknown. 
In the following we determine their {\it minimal} values when accommodating either a 3-flavour  or a 3 + 1 (or more) -flavour mixing schemes.
The different possibilities are summarised in Table~\ref{massspectrum}.

\begin{table}[htbp]
 \begin{tabular}{|c|c|c||c|c|c||c|c|}
\hline
\hspace*{-3mm}
\begin{tabular}{c}
\#   {\footnotesize new} \\
{\footnotesize fields}
\end{tabular} \hspace*{-3mm}
& $\# \nu_R$ & $\# s$ &
$  \# {m^2_i}^{(0)} =0$
&
$  \# {m^2_i}^{(1,2)} \neq0$
&\begin{tabular}{c}
\# {\footnotesize of} \\ {\footnotesize non-deg. }\\ {\footnotesize light} $m_i$\end{tabular} &  
\hspace*{-3mm} \begin{tabular}{c} {\footnotesize oscillation}  \\ {\footnotesize data:}\\ $\Delta m^2$ \end{tabular} \hspace*{-3mm} 
&  \hspace*{-3mm}
\begin{tabular}{c} {\footnotesize oscillation} \\ 
{\footnotesize data:}\\  $\Delta m^2$ \& $U_\text{\scriptsize PMNS} $\end{tabular} 
\hspace*{-3mm}
\\
\hline
2&1&1&3&1&2&\XSolidBrush &\XSolidBrush  \\
\hline
3&1&2&4&2&3&\Checkmark (s) &\XSolidBrush  \\
\hline
3&2&1&2&1 &2&\XSolidBrush &\XSolidBrush \\
\hline
4&1&3&5&3&4&\Checkmark (a)&\XSolidBrush \\
\hline
4&2&2&3&2&3&\Checkmark (s) &\Checkmark \\
\hline
4&3&1&1&1&1&\XSolidBrush &\XSolidBrush \\
\hline
5&2&3&4&3&4&\Checkmark (a) &\Checkmark \\
\hline
5&3&2&2&2&2&\XSolidBrush &\XSolidBrush \\
\hline
6&3&3&3&3&3&\Checkmark (s) &\Checkmark\\
\hline
 \end{tabular}
\caption{Tree-level neutrino mass spectra for different choices of the number of additional fields, $\nu_R$ and $s$, and different properties of the light neutrino spectrum (see text for details and for description of used symbols). We limit the table to the case where the maximum number of additional singlet fields is six.}
\label{massspectrum}
\end{table}

The first three columns of Table~\ref{massspectrum}   indicate 
the total number of additional  fermionic singlets $\# \nu_R+ \#s$,  $\# \nu_R$ and  $ \#s$, respectively. 
The fourth column contains the number of  massless eigenstates at zeroth order 
(in the absence of accidental cancellations between the a priori independent entries {of the mass matrix}). 
Always in the absence of accidental cancellations,
the fifth column displays how many  massless eigenstates acquire mass once higher order  corrections from perturbations  
are taken into account (see Appendix \ref{AppendixA}): although massive, these states remain light
since the corresponding masses are proportional to  entries of $m$ and $\mu$ 
(this can be inferred from the one generation illustrative case, see Eq.~(\ref{1genneutrinomass(0)})). 
It is important to notice that states which are already massive at zeroth order have masses proportional to the $d$ and $n$ matrix entries. Finally, the sixth column contains information on the number of non-degenerate light mass eigenstates predicted by each of the different ISS configurations considered.

The last two columns provide information on the phenomenological viability of the different ISS realisations. Firstly, 
neutrino oscillation experiments require at least two independent oscillation frequencies ($\Delta m_{ij}^{{2}}$); if 
there are less than 3 different light masses, the model is then excluded by observation, and this is denoted by a  \XSolidBrush\ .
Models with 3 different light masses can generate the correct neutrino mass spectrum and are marked with a \Checkmark~(s) 
in the seventh column of the table. 

Interestingly,  models with 4 different light masses could potentially explain the (anti)neutrino anomalies reported by the short baseline experiments LSND~\cite{Aguilar:2001ty} and MiniBooNE~\cite{miniboone:I}, the Gallium anomaly  in radioactive 
source experiments~\cite{gallium:I} and the reactor antineutrino anomalies~\cite{reactor:I}.
Such configurations, leading to a 3 + 1-mixing scheme (see for example~\cite{Giunti:2012tn}) are 
indicated by a \Checkmark~(a) in the seventh column of Table~\ref{massspectrum}. 

For all cases with a viable mass spectrum - either (s) or (a) - we have then verified if the observed mixing pattern could be successfully reproduced. Should this be the case,  a \Checkmark is present in the eighth column of the table. 

\medskip
As can be seen from the information summarised on Table~\ref{massspectrum}, the  
simplest model\footnote{In our study, the  first scenario ("(1,1) ISS")  would have corresponded to the $n=1$  scenario  in~\cite{Gavela:2009cd}, provided  the entry $(1,3)$ of Eq. (\ref{isszeromatrix}) was different from zero.} which could accommodate the observed neutrino spectrum is the one with $ \left( \# \nu_R=1, \# s=2 \right) $, which will be here denoted as "(1,2) ISS". 
It predicts $4$ light eigenstates, two of which are massive; provided that the latter are non-degenerate, one could have two independent mass squared differences (corresponding to the solar and atmospheric mass differences).  Notice however that this model cannot provide the observed leptonic mixing matrix $U_\text{PMNS}$. This is a consequence 
of having one of its light mass eigenstates dominated by sterile components, and as such it cannot be identified with a SM active neutrino. A similar problem is present for the "(1,3) ISS" configuration, which although in principle accommodating the correct neutrino mass spectrum fails to provide the observed mixings. 

The scenarios $ \left( \# \nu_R=2, \# s=1 \right) $ and $ \left( \# \nu_R=3, \# s=1 \right) $ could in principle accommodate neutrino data (masses and mixing) in the limiting case where sterile fields decouple, i.e. the matrix $n\to  0$ in Eq. (\ref{generalmatrix}). We further emphasise here that we are not in this situation (of a type I seesaw with 2 or 3 right-handed neutrinos), and these two scenarios do not comply with neutrino data. In the case of  $ \left( \# \nu_R=2, \# s=1 \right) $,  the corresponding mass spectrum contains one massless active neutrino, one light active while the third active one is too heavy to explain solar and atmospheric oscillation frequencies. 
A similar situation occurs for the $ \left( \# \nu_R=3, \# s=1 \right) $ case, where one has only one light active neutrino and two (too) heavy active ones.  

From this simple analysis and in view of  Table~\ref{massspectrum}, the first realisation  of the inverse seesaw (with $\#s\ne 0$) possibly accommodating neutrino data is $ \left( \# \nu_R=2, \# s=2 \right) $, which we define to be the minimal one under the previous assumption of Eq. (\ref{condition}), 
 hereafter denoted by "(2,2)  ISS" realisation. Notice that this solution  corresponds to the minimal model found in~\cite{Gavela:2009cd} in the case where no lepton number violating Yukawa couplings are allowed. 
This "(2,2)  ISS" scenario  does not provide an explanation for the reactor anomaly;  the next (to minimal) ISS realisation 
which could explain such anomaly is the one with $ \left( \# \nu_R=2, \# s=3 \right) $,
 which we denote  by "(2,3) ISS" configuration.

Before addressing in detail the phenomenology of each minimal framework above identified, we will briefly comment 
on some aspects intrinsic to all ISS realisations.

\subsection{Different neutrino mass scales}\label{generaliss}

As a function of the number of generations for the sterile fields ($\#s\ne 0, \#\nu_R\ne 0$), 
the model always exhibits $\#\nu_L + (\#s-\#\nu_R)$ light mass eigenstates. These states would be massless at zeroth order, and their masses arise from higher order corrections (in perturbation) due to the block-diagonal matrix which now generalizes $\Delta M$, see Eq.~(\ref{DeltaM}).
In addition, the full spectrum contains heavy states with masses  $\sim \mathcal{O}(n_{i,j})+\mathcal{O}(d_{i,j})$, which form 
$\#\nu_R$ pseudo-Dirac pairs with mass differences $\sim \mathcal{O}(\mu_{i,j}), \ \mathcal{O}(m_{i,j})$. 
In the limit where lepton number is conserved (i.e. $\Delta M=0$) these states become Dirac particles.

The low-energy phenomenology of these models is determined by two quantities: the scale of the Lepton Number Violating  parameters $\mu$ and  the ratio between the scale of the Dirac mass terms $d$ and that of the $n$ mass matrix, denoted by $k$.
To understand the key r\^ole of these  quantities, let us consider again the illustrative one-generation model 
(i.e. $\# \nu_L = \# \nu_R = \# s =1$) of 
Section~\ref{largeM}. The active neutrino mass of Eq.~(\ref{1genneutrinomass(0)}) can be rewritten as
$m_\nu = |\mu| k^2/(1+k^2)$, with $k=|d|/|n|$. 
In the realistic case of several generations, $d,n,\mu$ are matrices, and  these considerations loosely apply to the order of magnitude of their entries.
The ratio $k$ is directly related to  deviations from unitarity of the leptonic mixing matrix, as shown in Appendix \ref{AppendixA}, Eq.~(\ref{1genneutrinoeigenvec}).
Constraints on the non-unitarity of the PMNS matrix impose that $k$ should not be too large; 
as we will discuss in the section devoted to the numerical analysis,  
solutions in agreement with experimental data can be found if, and only if,
$\mathcal{O}(d)/\mathcal{O}(n) \lesssim 10^{-1}$. 
These features are shared by the different realistic extensions presented in Table~\ref{massspectrum}.
 
 The mass spectrum of the ISS models is thus characterised by either 2 or 3 different mass scales (as illustrated in Fig.~\ref{frize}):

\begin{figure}[htbp]
 \begin{center}
  \includegraphics[width=10cm]{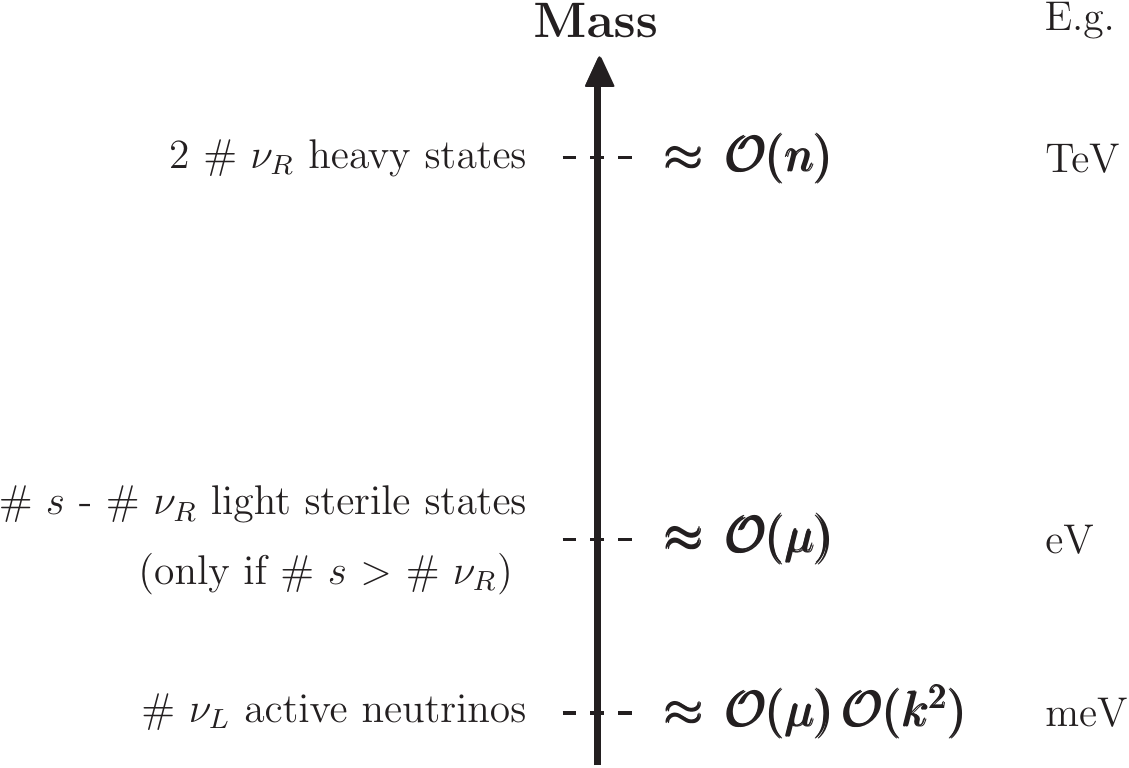}
\caption{Pictorial representation of typical scales for the neutrino mass spectrum in several ISS realisations.}
 \label{frize}
 \end{center}
\end{figure}

\begin{itemize}
\item the one of the light active neutrinos  $\sim \mathcal{O}(\mu) \mathcal{O}(k^2)$;
\item the scale corresponding to the heavy states, roughly $\mathcal{O}(d)+\mathcal{O}(n)\approx \mathcal{O}(n)$;
\item an intermediate scale of order $\mathcal{O}(\mu)$  corresponding to $\#s-\#\nu_R$ sterile light states (only present when $\#s>\#\nu_R$).

\end{itemize}

\subsection{Removing unphysical parameters}

The relevant leptonic terms of a general inverse seesaw  Lagrangian can be written in the following compact  form, 
\be \label{lag22}
\mathcal{L}_\text{leptonic} \,= \,\mathcal{L}_\text{kinetic} + \mathcal{L}_\text{mass} + 
\mathcal{L}_\text{CC}+ \mathcal{L}_\text{NC}+ \mathcal{L}_\text{em}\,,
\ee
where
\begin{eqnarray}
\mathcal{L}_\text{kinetic}&=&  i \,\overline{e_L}^\alpha \,\slashed{\partial} \,
\delta_{\alpha,\beta} \,e_L^\beta + 
i \,\overline{e_R}^\alpha \,\slashed{\partial} \,\delta_{\alpha,\beta} \,e_R^\beta + 
i \,\overline{\nu_L}^\alpha \,\slashed{\partial} \,\delta_{\alpha,\beta} \,\nu_L^\beta+ 
i \,\overline{\nu_R}^i \,\slashed{\partial} \,\delta_{i,j} \,\nu_R^j + 
i \,\overline{s}^a \,\slashed{\partial} \,\delta_{a,b} \,s^b\,, \nonumber \\
\mathcal{L}_\text{mass}&=& 
- \overline{e_R}^\alpha \,{\mathfrak{m}}_{\alpha,\beta} \,e_L^\beta - 
\overline{\nu_R}^i \,d^T_{i,\alpha} \,\nu_L^\alpha -  
\overline{\nu_R}^i \,m_{i,j} \, {\nu_R^c}^j - 
\overline{\nu_R}^i \,n_{i,a} \,s^a - 
\overline{s^c}^a \,\mu_{a, b} \,s^b + \text{h.c.} \,, \nonumber \\
 \mathcal{L}_\text{CC}&=& 
 \frac{g}{\sqrt{2}} \,\overline{e_L}^\alpha \,\slashed{W}^- \,\delta_{\alpha,\beta} \,\nu_L^\beta + \text{h.c.}\,, 
 \nonumber \\
\mathcal{L}_\text{NC} &=& 
\frac{g}{\cos{\theta_W}} \,\left\{ \frac{1}{2} \,\left[ \overline{\nu_L}^\alpha \,\gamma_\mu \,
\delta_{\alpha,\beta}  \,\nu_L^\beta - \overline{e_L}^\alpha \,\gamma_\mu \,
\delta_{\alpha,\beta}\, e_L^\beta \right] - \sin^2{\theta_W} \,J_\mu^\text{em} \right\} 
\,Z^\mu\,, \nonumber \\
\mathcal{L}_\text{em} &=& e \,J_\mu^\text{em} \,A^\mu.
\end{eqnarray}
In the above equation  $\alpha,\beta=1,2,3$, $i,j=1,\dots,\# \nu_R$ and $a,b=1,\dots,\# s$. 
The total number  $n_u$ of physical and non-physical parameters in the mass matrices present in the  
Lagrangian of Eq.~(\ref{lag22}) is equal to 
\be
n_u=18 + 6\ \# \nu_R + \# \nu_R (\# \nu_R + 1) + \# s (\#s + 1) + 2\ \# \nu_R\ \# s\,,
\ee
and detailed in Table~\ref{numorder22}.

\begin{table}[h!]
\begin{center}
\begin{tabular}{|c|c|}
\hline
{\footnotesize Matrix} & {\footnotesize Total number of parameters} \\
\hline
${\mathfrak{m}}$ & $18$ \\
$d$ & $6\times \# \nu_R$\\
$n$ & $2\times \# \nu_R\times\# s$ \\
$m$ & $\# \nu_R\times (\# \nu_R + 1)$ \\
$\mu$ & $\# s \times(\#s + 1)$ \\
\hline
{\footnotesize Total} & $18 + \# \nu_R (7 + \# \nu_R +2\ \# s) + \# s (\# s+1)$\\
\hline
\end{tabular}
\end{center}
\caption{Total number of physical and non-physical parameters in the Lagrangian of Eq.~(\ref{lag22}).}
\label{numorder22}
\end{table}

In order to determine the actual number of  physical parameters, one has to identify all  independent transformations under which the Lagrangian of Eq.~(\ref{lag22}) is invariant. One finds four classes of transformations with the following unitary matrices: 
\begin{enumerate}
\item $U^{L}$ ($3 \times 3$):
\noindent\be\label{sym1.22} \hspace*{-8mm}
e_L^\alpha \rightarrow  U^L_{\alpha,\beta}e_L^\beta\,,
\quad
{\mathfrak{m}}_{\alpha,\beta} \rightarrow {\mathfrak{m}}_{\alpha,\gamma} 
{U^L}^\dagger_{\gamma,\beta}\,,
\quad
\nu_L^\alpha \rightarrow U^L_{\alpha,\beta} \nu_L^\beta\,,
\quad
d^T_{i,\alpha} \rightarrow d^T_{i,\beta} {U^L}^\dagger_{\beta,\alpha}\,;
\ee

\item  $U^{R}$ ($3 \times 3$):
\be\label{sym2.22} \hspace*{-8mm}
e_R^\alpha \rightarrow U_{\alpha,\beta}^R e_R^\beta\,,
\quad
{\mathfrak{m}}_{\alpha,\beta} \rightarrow U^R_{\alpha,\gamma} {\mathfrak{m}}_{\gamma,\beta}\,;
\ee

\item $U^{\nu_R}$ ($\# \nu_R \times \# \nu_R$):
\be\label{sym3.22} \hspace*{-8mm}
{\nu_R^c}^i  \rightarrow  U^{\nu_R}_{i,j} {\nu_R^c}^j\,,
\quad
m_{i,j}  \rightarrow  {U^{\nu_R}}^*_{i, k} m_{k,l} {U^{\nu_R}}^\dagger_{l,j}\,,
\quad
d^T_{i,\alpha}  \rightarrow  {U^{\nu_R}}^*_{i, j} d_{j,\alpha}^T\,,
\quad
n_{i,a}  \rightarrow  {U^{\nu_R}}^*_{i, j} n_{j,a}\,;
\ee

\item $U^s$ ($\# s \times \# s$):
\be\label{sym4.22} \hspace*{-8mm}
s^a  \rightarrow  U^s_{a,b} s^b\,,
\quad
\mu_{a,b}  \rightarrow {U^s}^*_{a, c} \mu_{c,d} {U^s}^\dagger_{d, b}\,,
\quad
n_{i,a} \rightarrow  n_{i, b} {U^s}^\dagger_{b,a}\,.
\ee

\end{enumerate}
The number of parameters defining the transformations of Eqs.~(\ref{sym1.22} - \ref{sym4.22}) 
is $n_t=18 +(\# \nu_R)^2 + (\# s)^2$, as shown in Table~\ref{numparsym22}, so that the 
number of physical parameters $n_p$ thus reduces to 
\be \label{nphys}
n_p\,=\,n_u-n_t\,=\,7\ \# \nu_R +\# s + 2 \# \nu_R \ \# s\,.
\ee

\begin{table}[h!]
\begin{center}
\begin{tabular}{|l|c|}
\hline 
{\footnotesize Matrix} & {\footnotesize Number of free parameters}\\
\hline
$U^L$ & $9$ \\
$U^R$ & $9$ \\
$ U^{\nu_R}$ & $(\# \nu_R)^2$ \\
$U^s$ & $(\# s)^2$\\
\hline 
{\footnotesize Total} & $18 +(\# \nu_R)^2 + (\# s)^2$\\
\hline
\end{tabular}
\end{center}
\caption{Number of parameters defining the transformations of Eqs.~(\ref{sym1.22} - \ref{sym4.22}).}
\label{numparsym22}
\end{table}

Since $\mathcal{L}_\text{kin}$ is invariant under 
 each of the transformations of  Eqs.~(\ref{sym1.22} - \ref{sym4.22}), we can use 
 the latter to redefine the fields and cast the  mass matrices only in terms of physical parameters. 
 For instance,  with the transformations of Eqs.~(\ref{sym1.22}, \ref{sym2.22}), one  can  choose a basis in which the charged leptonic matrix ${\mathfrak{m}}$ is real and diagonal,  and similarly for the symmetric Majorana mass matrices $m$ and $\mu$ (in this case using Eqs.~(\ref{sym3.22}, \ref{sym4.22})). 
Finally,  one can eliminate three phases from the Dirac mass matrix $d$  while  keeping ${\mathfrak{m}}$ real, 
via a combination of transformations of Eq.~(\ref{sym1.22}) and Eq.~(\ref{sym2.22}).
In this simple example, there are exactly $n_p$ free parameters, as summarised in Table~\ref{physpar}.

\begin{table}[h]
\begin{center}
\begin{tabular}{|c|c|c|c|}
\hline
{\footnotesize Matrix} & {\footnotesize \# of moduli} & {\footnotesize \# of phases} &  
{\footnotesize Total} \\
\hline
{\footnotesize Diagonal and real ${\mathfrak{m}}$ }& $3$ & $0$ & $3$ \\
{\footnotesize $d$ with three real entries} & $3\ \# \nu_R$ & $3\ \# \nu_R-3$ & $6\ \# \nu_R-3$ \\
{\footnotesize Real and diagonal} $m$ & $\# \nu_R$ & $0$ & $\# \nu_R$ \\
 $n$ & $\# \nu_R \ \# s$ & $\# \nu_R \ \# s$ & $2 \ \# \nu_R \ \# s$ \\
{\footnotesize Real and diagonal $\mu$} & $\# s$ & $0$ & $\# s$\\
\hline
{\footnotesize Total} & \multicolumn{3}{ |c| }{$7\ \# \nu_R +\# s + 2 \# \nu_R \ \# s$}\\
\hline 
\end{tabular}
\end{center}
\caption{Example of a basis in which all unphysical degrees of freedom have been rotated away. 
}
\label{physpar}
\end{table}

\section{Effects of fermionic gauge singlets and constraints on the ISS parameters}\label{Sec:constrains}

In addition to succeeding in accommodating neutrino oscillation data, models with sterile fermions 
are severely constrained, since the mixings of the sterile neutrinos with the active
left-handed states might induce contributions to several observables, leading to conflict with experimental 
data. The mixings of the sterile neutrinos with the active
left-handed states  imply a departure from unitarity of the $3\times 3$ 
$U_\text{PMNS}$ matrix, which 
can have an impact on several observables, inducing deviations from the SM predictions.
Bounds on the non-unitarity of the PMNS were derived in~\cite{Antusch:2008fk}, using 
Non-Standard Interactions (NSI). These bounds are especially relevant in our analysis when 
the masses of the sterile states are heavier than the GeV, but some are still lighter than 174 GeV.

If the sterile states are sufficiently light and  have large mixings with the active neutrinos
(as for example in the inverse seesaw~\cite{Mohapatra:1986bd}
, 
the $\nu$SM \cite{Asaka:2005an} and the 
low-scale  type I seesaw~\cite{Gavela:2009cd,Ibarra:2010xw,lowscale.typeI}), 
then the deviations from unitarity  
of the PMNS mixing matrix can be sizeable, and lead to 
(tree-level) corrections to the $W \ell \nu$ vertex. This will have a significant impact to 
several observables, such as corrections to the invisible $Z$ decay width~\cite{Akhmedov:2013hec}, 
significant contributions to lepton flavour 
universality (LFU) violation observables~\cite{Shrock, Abada:2012mc, Abada:2013aba}, 
and new contributions to numerous low-energy rare decays.

Another important  constraint concerns charged lepton flavour violation (cLFV)  
since the modified $W \ell \nu$ vertex gives rise to cLFV processes, typically at rates higher than the current bounds unless the active-sterile mixings are  small~\cite{Mohapatra:1986bd,GonzalezGarcia:1988rw,Deppisch:2004fa,Ilakovac:1994kj}.  
In the case of $\mu \to e \gamma$ decays, the rate induced by the presence of the sterile states is given by~\cite{muegamma}:
\be \label{muegamma}
\text{Br}(\mu \to e \gamma) = \frac{3 \alpha_{\text{em}}}{32 \pi}\left|\sum_{i} U_{\mu i}^* U_{e i} G\left(\frac{m_i^2}{M_W^2}\right)\right|^2 ,
\ee
where the index $i$ runs over all neutrino states,  $U$ is 
the leptonic mixing matrix obtained after diagonalization of the mass matrix and $G$ is the associated loop function.
The current bound on this branching ratio is $\text{Br}(\mu \to e \gamma) < 5.7 \times 10^{-13}$ at 90\% C.L., as reported 
very recently by the MEG experiment~\cite{Adam:2013mnn}. This will prove to be 
the most relevant LFV bound in most of our scenarios with light sterile neutrinos.
\vskip 0.5cm
Constraints arising from  neutrinoless double beta ($0\nu 2 \beta$) decay bounds can be particularly relevant, 
since in the ISS  the heavy sterile states also contribute to the process.
The effective neutrino mass $m^{\nu_e}_{\text{eff}}$, to which  the amplitude 
of the $0\nu 2 \beta$ process is proportional,  can receive further corrections with respect to the standard expression, 
$\sum_{i=1}^3 U_{e,i}^2 m_{\nu_i}$. 
Since the heavy Majorana states mix to form pairs of pseudo-Dirac states, their contribution is proportional to their mass difference weighted by the $\nu_e$-sterile mixing.  Each Majorana state thus contributes 
to the amplitude of a $0\nu \beta \beta$ decay as~\cite{Blennow:2010th} 
\be 
A_i \propto m_i U_{e,i}^2 M^{0\nu \beta \beta}(m_i)\,,
\ee
where $M^{0\nu \beta \beta}(m_i)$ is the nuclear matrix element that characterises 
the process. The latter is a function of the neutrino mass $m_i$ and depends on the nucleus that 
undergoes the $0\nu \beta \beta$ transition. It can be satisfactorily approximated by the analytic expression
\be 
M^{0\nu \beta \beta}(m_i) \simeq M^{0\nu \beta \beta}(0) \frac{p^2}{p^2-m_i^2},
\ee
where $p^2 \simeq - (125 \mbox{ MeV})^2$ is the virtual momentum of the neutrino. 
We will conduct a detailed analysis of the impact of two minimal  ISS  realisations, the "(2,2) ISS" and "(2,3) ISS",  
on the effective electron neutrino mass in Sections~\ref{nu0bb} and \ref{nu0bb-bis}.

\medskip
Moreover, if the typical scale of the new states is sufficiently light, they can be produced in collider or low-energy experiments, thus being subject to further constraints~\cite{BhupalDev:2012zg}.
Robust laboratory bounds arise from direct searches for sterile neutrinos, which can be produced in meson decays such as $\pi^\pm \to \mu^\pm \nu$, with rates that depend on their mixing with the active neutrinos. Therefore, negative searches for monochromatic lines in the muon spectrum can be translated into bounds on the active-sterile 
mixing~\cite{Kusenko:2009up,Atre:2009rg}.
\medskip

All the above mentioned bounds will be taken into account in our subsequent numerical analysis of the two minimal ISS realisations.

\section{Phenomenological analysis}\label{Sec:analysis}

Although it is possible to derive analytical expressions for the neutrino mass eigenvalues and leptonic mixing matrix (see Appendices~\ref{AppendixA} and \ref{AppendixB}), 
these expressions are  lengthy and involved, and do not easily convey the general features and behaviour of the ISS configurations investigated. 
We thus conduct a numerical analysis for each of the minimal "(2,2) ISS" and "(2,3) ISS" realisations. 
In order to unveil some interesting features, we performed a scan of the parameter space (corresponding 
to all the entries of the mass matrix; in our analysis we will not address the effect of CP violating phases, both Dirac and Majorana). This also allows to numerically compute interesting quantities, as for instance
the effective mass in $0\nu2\beta$ decay amplitude.
All the  constraints listed in Section~\ref{Sec:constrains} were implemented. We proceed to discuss the results
in the following sections.

\subsection{The "(2,2) ISS" realisation}\label{22realisation}
Some aspects of this model have already been studied, in particular CP violation and Non Standard 
Interactions~\cite{malinsky}. 
We have determined the neutrino spectrum and the leptonic mixing matrix using  a perturbative approach, whose details are summarised in Appendix~\ref{AppendixB}.
At second order in the perturbative expansion, the light neutrino  spectrum is given by:
\be\label{22issmasses-t}
{m_1^2}^{(2)}\,=\,0,\quad
{m_2^2}^{(2)}\,=\,\frac{b-\sqrt{b^2+4 c}}{2},\quad
{m_3^2}^{(2)}\,=\,\frac{b+\sqrt{b^2+4 c}}{2}\,,
\ee
where the parameters $b$ and $c$ are defined in terms of the entries of the (2,2) mass matrix; these expressions are 
lengthy, as explained in Appendix~\ref{AppendixB}. 
Notice that $b$ and $c$ do not depend on the submatrix $m$ of the mass matrix of Eq.~(\ref{generalmatrix}).

Having one massless eigenstate (to all orders in perturbation theory) is a feature of this minimal 
"(2,2) ISS" model (see also Table~\ref{massspectrum}). 
The expressions of Eq.~(\ref{22issmasses-t}) 
allow to easily understand why the "(2,2) ISS" model strongly prefers the normal 
hierarchy scheme. 
In order to accommodate an inverted hierarchy, i.e. $m_2^2 \simeq m_3^2 \simeq 10^{-3}\ \text{eV}^2$ and 
$m_3^2  - m_2^2 \simeq 10^{-5}\ \text{eV}^2$, one would be led  to comply with 
$10^{-6} \text{ eV}^4 +4c \simeq 10^{-10} \text{ eV}^4$. This amounts to an extreme fine-tuning.  
Although some solutions can indeed be found (see the numerical studies of the following section), 
it should be stressed that accommodating a NH spectrum also requires a certain amount of fine-tuning.

Even if useful when addressing  the issue of the hierarchy of the light neutrino spectrum, the analytical expressions we have  
derived for the neutrino masses and leptonic mixings cannot be used to extract general features, nor to 
infer the  magnitude of the fundamental scales of the 
ISS model (i.e. the magnitude of the entries of the matrices $\mu$, $m$, ...).
To do so, we performed numerical scans of the "(2,2) ISS"  parameter space, the result of which we proceed to report. 

\subsubsection{Mass hierarchy}

As discussed in Section~\ref{Sec:towards} and illustrated in Fig.~\ref{frize},  the low-energy phenomenology of a "(2,2) ISS" model is 
determined by two scales: 
that of the LNV parameter $\mu$, and the ratio $k$ between the magnitude of the entries of the $d$ and $n$ matrices, 
see Appendix~\ref{AppendixA}.

 In our numerical analysis, we randomly scan over all parameters: the entries of  the $d$ and $n$ submatrices are varied such that the obtained mixing matrix $U_\text{PMNS} $ is in agreement with oscillation data (global fits to both hierarchies, normal and inverted~\cite{GonzalezGarcia:2012sz}) and the interval of variation for the entries of $\mu$ is chosen to ensure that the largest neutrino squared mass value $ \sim 2.4 \times 10^{-3} \text{ eV}^2$. 
 While scanning over the parameter space, we always make sure that Eq. (\ref{condition}) is fulfilled,  assuming $\mu$ and $m$ to be  of the same order of magnitude. Moreover,  we take all parameters to be real (leading to vanishing Dirac and Majorana phases, and hence no contributions to leptonic electric dipole moments). 

\begin{figure}[htbp]
 \begin{tabular}{ll}
\includegraphics[width=75mm]{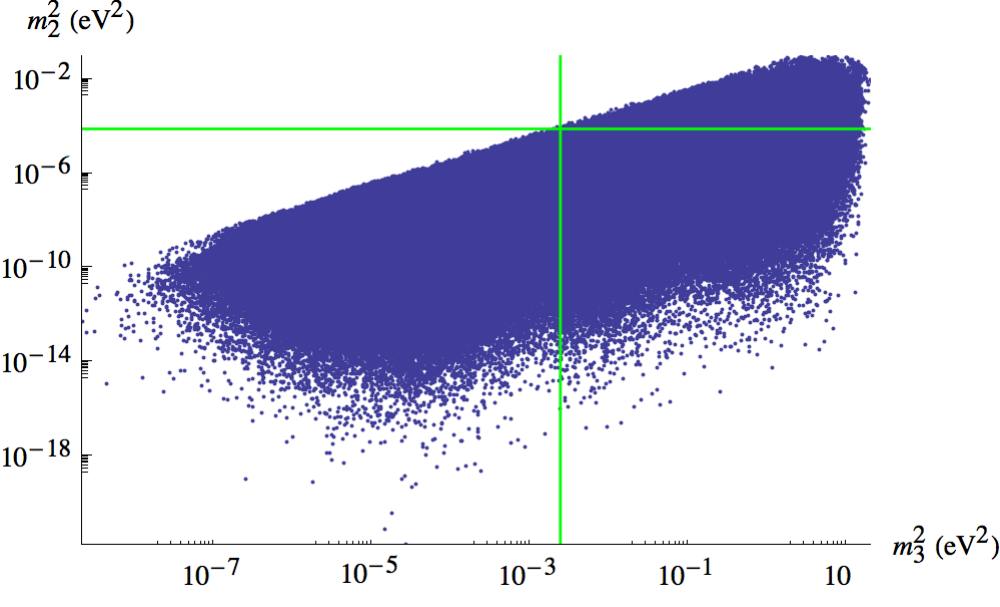}
\hspace*{2mm}&\hspace*{2mm}
\includegraphics[width=75mm]{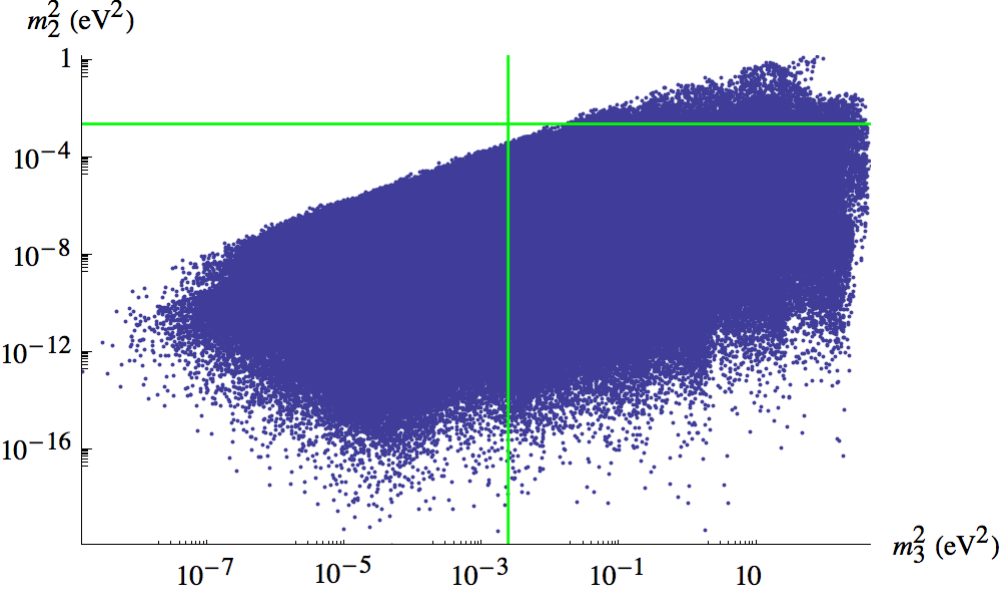}
\end{tabular}
\caption{Squared masses of the active neutrinos for the "(2,2) ISS" model (the lightest neutrino is massless). All points displayed fulfill the experimental constraints on the PMNS entries for the NH (left) and IH (right) schemes. The green lines denote the experimental best fit values \cite{GonzalezGarcia:2012sz} in the NH or IH schemes.
 The scan details are summarised in the text.}
\label{m1m2sq1.22}
\end{figure}

In Figure~\ref{m1m2sq1.22}, 
we collect the  values of the squared masses  $ m_{i}^2$ imposing that all the obtained mixing angles $\theta_{ij}$  are in agreement with oscillation data (for both cases of hierarchy, NH and IH).  
Leading to this figure, we varied for the left (right) panel the entries of each submatrix (see Eq.~(\ref{22massmatrix})) as $d_{i,j} \in  [10^{6},10^{8}] \text{ eV}$, $n_{i,j} \in  [10^{7},10^{9}] \text{ eV}$ ($n_{i,j} \in  [10^{8},10^{10}] \text{ eV}$), and  $m_{i,j},\mu_{i,j} \in  [10^{-3},10] \text{ eV}$ ($m_{i,j},\mu_{i,j} \in  [10^{-1},10^2] \text{ eV}$).

The best fit values for the mass eigenvalues resulting from the global analysis of the oscillation experiments~\cite{GonzalezGarcia:2012sz} are indicated in Fig.~\ref{m1m2sq1.22} by horizontal and vertical lines. 
This example clearly illustrates the analytical result found in Section \ref{22realisation} (as well as in Appendix \ref{AppendixB}):  
the "(2,2) ISS" model favours a normal hierarchical scheme - the inverted hierarchy requiring in this case an extreme fine tuning of the parameters, see Eq. (\ref{22issmasses-t}).  This  can be seen on the right panel of Fig.~\ref{m1m2sq1.22},  as no solutions can be encountered for an IH scheme (corresponding to 
$\Delta{m^2_{32}}\sim 10^{-5} \text{eV}^2$ together with $m_2^2\sim m_3^2\sim 10^{-3} \text{eV}^2$). 
Moreover, as can be seen on the left panel of Fig.~\ref{m1m2sq1.22}, for the NH scheme, finding solutions for the light neutrino masses in agreement with data is possible although difficult.

\subsubsection{Constraints from unitarity}\label{NSI}
The non-observation of NSI in the leptonic sector as induced by the deviation from unitarity of the $U_\text{PMNS}$ matrix due to the presence of additional fermions puts stringent constraints~\cite{Antusch:2008fk} on the ISS parameter space.

The non-unitarity effects can be quantified by 
\be\label{nsi}
\epsilon_{\alpha \beta}\equiv \left| \sum_{i=4}^7 U_{\alpha,i} \,U_{i,\beta}^\dagger \right|\,=\,\left|\delta_{\alpha,\beta} - \left(N\,N^\dagger\right)_{\alpha,\beta}\right|\,, 
\ee
where $N$ is the  $3 \times 3$ submatrix encoding the  mixings between the active neutrinos and  the charged leptons, i.e. the PMNS matrix. 
Depending on the mass regime for the sterile fermions (above or below the EW scale)  the constraints on $ \left(NN^\dagger\right)$ are different~\cite{Antusch:2008fk}.
We thus identify the following mass regimes for our sample of "(2,2) ISS"  mass matrices:
\begin{itemize}
\item no (or only some) sterile states are above 1 GeV - implying that not all the extra states can be indeed integrated out; 
the NSI constraints of~\cite{Antusch:2008fk} do not apply in this case;
\item all sterile states are heavier than 1 GeV, but do not necessarily lie above the EW scale, $\Lambda_\text{EW} \sim$174 GeV; 

\item all sterile states are heavier than $\Lambda_\text{EW}$.

\end{itemize}
When appropriate, we thus compute the amount  of non-unitarity from Eq.~(\ref{nsi}), and apply the corresponding bounds, to further constrain the ISS parameter space.

Notice that in the ISS models the non-unitarity effects are proportional to the ratio $\mathcal{O}(d)/\mathcal{O}(n)$ (see for example the neutrino mass eigenvector expression for the one-generation model 
(Eq.~\ref{1genneutrinoeigenvec})).

\begin{figure}[htbp]
 \begin{tabular}{ll}
\includegraphics[width=75mm]{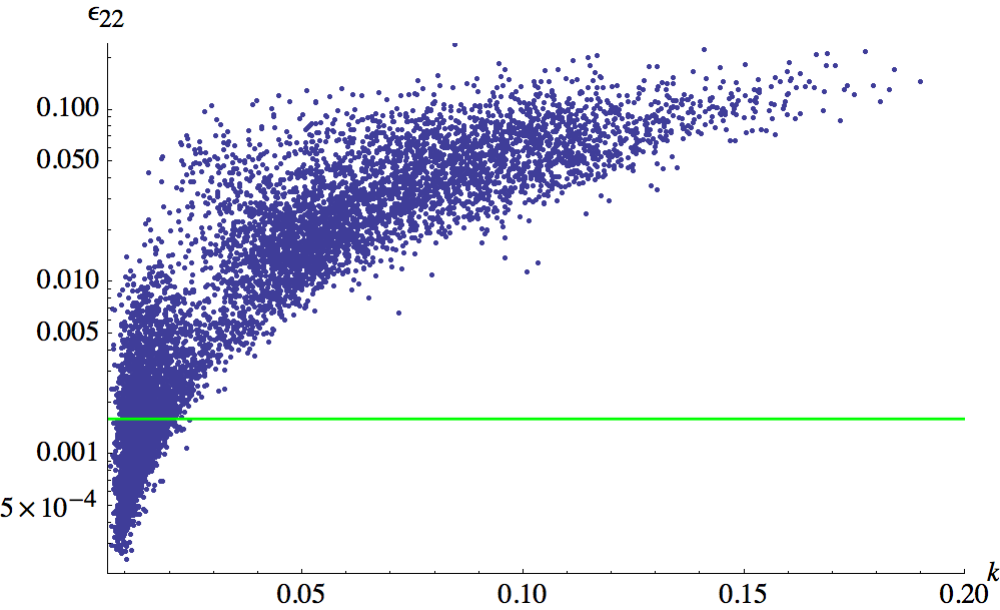}
\hspace*{2mm}&\hspace*{2mm}
\includegraphics[width=75mm]{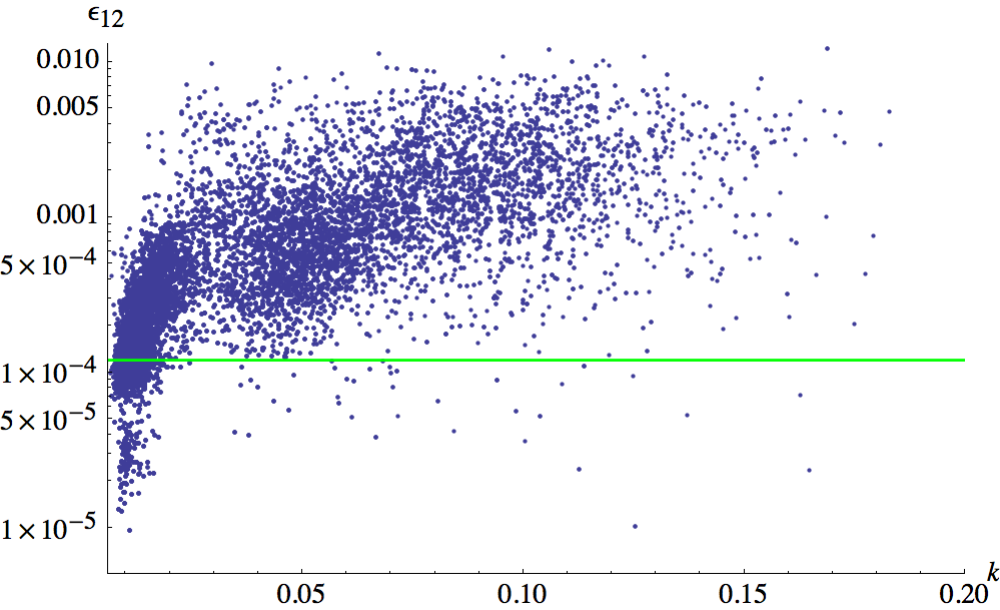}\end{tabular}
\caption{Examples of $  \epsilon =\left|\mathbf{1}- \left(N\,N^\dagger\right)\right|$ entries, as  a function of an effective factor $k$ (see Eq.~(\ref{k})).
On the left, $ \epsilon_{22}$, for a  
mass regime in which the sterile neutrino masses are  between 1 GeV and $\Lambda_\text{EW}$; on the right, $ \epsilon_{12}$, in the regime where all sterile states are heavier than  $\Lambda_\text{EW}$.
The green lines indicate the corresponding upper bounds~\cite{Antusch:2008fk}. All points comply with oscillation data in the NH scheme.
 The scan details are summarised in the text.}
\label{nsi22gev}
\end{figure}

We display on Fig.~\ref{nsi22gev} the most constraining deviations from unitarity parametrised by 
$ \epsilon =| {\mathbf{1}- }\left(NN^\dagger\right)|$, see Eq.~(\ref{nsi}), as a function of  an effective factor  $k$ 
 generalizing the one  introduced in Section~\ref{generaliss}, which is defined as (see Eq.~(\ref{22massmatrix}) in Appendix~\ref{AppendixB}):
\be\label{k}
k\,=\,\frac{\left(d_{1,1}+d_{2,1}+d_{3,1}+d_{1,2}+d_{2,2}+d_{3,2}\right)/6}{\left( n_{1,1}+n_{2,2}\right)/2}\,.
\ee

Each point is generated with random values for the entries of the $d,n$ submatrices 
- but allowing the entries of each submatrix to vary at most over two orders of magnitude -,
 and such that the mass matrix would generate a PMNS matrix and a neutrino mass spectrum in agreement with experimental constraints (in the NH scheme). 
Leading to this figure (left  and right panels) , we varied  the entries of each submatrix (see Eq.~(\ref{22massmatrix})) as $d_{i,j} \in  [10^{3}, 1.7 \times 10^{11}] \text{ eV}$, $n_{i,j} \in  [5.5 \times  10^4,1.6 \times 10^{13}] \text{ eV}$ and  $m_{i,j},\mu_{i,j} \in  [5 \times 10^{-6}, 100] \text{ eV}$.

As can be seen from both panels of Fig.~\ref{nsi22gev}, NSI constraints significantly reduce the number of 
otherwise phenomenologically viable solutions for the "(2,2) ISS" model. 

\subsubsection{LFV constraints: Br(${\mu \to e \gamma}$)}
The presence of sterile fermions  may impact several observables in particular   LFV processes, with rates potentially larger than current bounds. We focus here on the radiative muon decay $\mu
\to e \gamma$, searched for by the MEG experiment~\cite{Adam:2013mnn} and which 
provides the most stringent constraint on the branching ratio of Eq.~(\ref{muegamma}).

\begin{figure}[htbp]
 \begin{center}
\includegraphics[width=8.5cm]{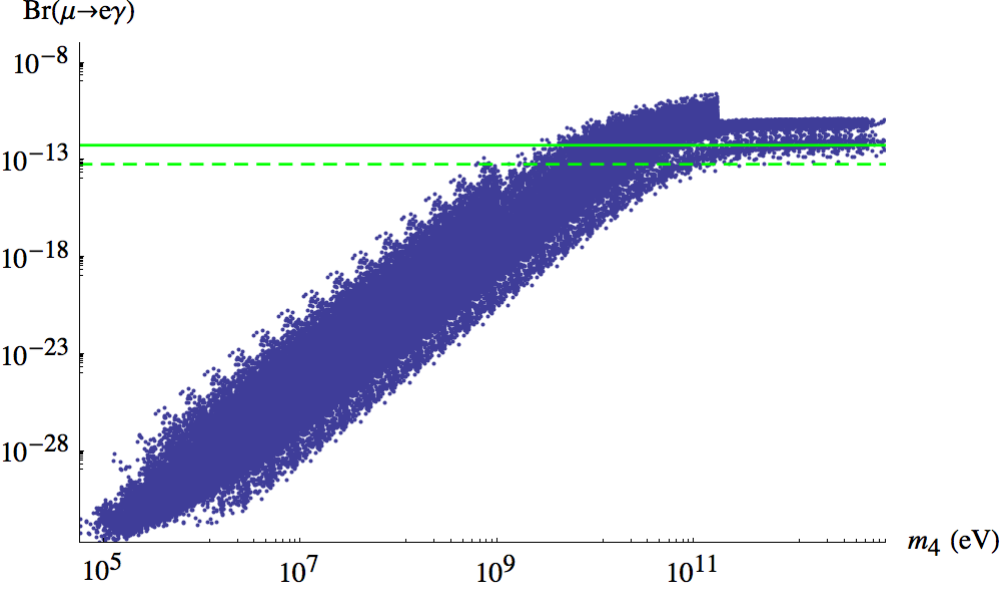}
\caption{Br(${\mu \to e \gamma}$) as a function of the mass of the lightest sterile state, $m_4$. 
The green full (dashed) horizontal lines denote MEG's current upper bound~\cite{Adam:2013mnn} (future sensitivity~\cite{Baldini:2013ke}). All points comply with oscillation data in the NH scheme and unitarity constraints. Scan details as in Fig.~\ref{nsi22gev}.}
\label{muegamma22}
 \end{center}
\end{figure}

 In Fig.~\ref{muegamma22}, we display this observable as a function of the mass of the lightest sterile state, $m_4$. The investigated parameter space (the same as the one leading to the previous figures) leads to contributions typically below the future experimental sensitivity. However, for $m_4$  heavier than $\sim 1$ GeV, one might observe a cLFV signal of the "(2,2) ISS" at MEG.

\subsubsection{Lepton number violating parameter space}
From the numerous numerical scans we conducted, certain features of the "(2,2) ISS" model became apparent: 
\begin{itemize} 
\item Low-energy neutrino data (i.e. masses and mixings)
can be accommodated if the entries in each of the submatrices of Eq.~(\ref{generalmatrix}) are allowed 
a strong  hierarchy  - varying at least over 2 orders of magnitude.
\item The model leads to a strongly hierarchical light neutrino mass spectrum, with   the second lightest neutrino mass being strongly suppressed with respect to the heaviest one (the first state being massless). 
\end{itemize}

The size of the LNV parameters (i.e. the entries of the $\mu$ submatrix - recall that the LNV matrix $m$ does not enter in the expression for the lightest neutrino mass eigenvalues, 
as derived in a perturbative approach - see for instance, Eq.~(\ref{1genneutrinomass(0)})) is bounded from below by PMNS matrix constraints, and from above by the naturalness requirement.
The lower limit is due to the fact that, to a good approximation, the entries of $d$ must be at least one order of magnitude smaller than those of  $n$ (in order to accommodate oscillation data).
In order to fulfill solar and atmospheric mass squared differences, and given that one typically has 
$k<10^{-1}$ (see Eq.~({\ref{1genneutrinomass(0)})), it follows that
\be 
|\mu| \ \gtrsim \ k^{-2}\times  8\times 10^{-3} \text{ eV} \ \gtrsim \ 8\times 10^{-1} \text{ eV}\,.
\ee

We have checked that the latter condition is indeed valid in the "(2,2) ISS" model;  
the lower values for the $\mu$ submatrix entries, 
in agreement with both $U_\text{PMNS}$ data  and neutrino mass squared differences 
are:
$\min|\mu_{i,i}| \sim 0.13$ eV, $\min|\mu_{i\neq j}| \sim 5 \times 10^{-6}$ eV.
The upper bound on the LNV parameters comes from  't Hooft naturalness criterium, even though
a clear definition regarding the naturalness of a small dimensionful parameter breaking some SM accidental 
symmetries does not exist. 
In this study, we have posited a "naturalness" upper limit of 100 eV on the entries of the submatrix $\mu$. 
This translates into a lower bound on the factor $k$ ({since $m_\nu \approx k^2 \mu$}).
\subsubsection{Neutrinoless double beta decay}\label{nu0bb}
When applied to the "(2,2) ISS" model, the effective neutrino mass $m^{\nu_e}_\text{eff}\ $ determining the 
amplitude of the neutrinoless double beta decay rate is given by (see Section~\ref{Sec:constrains})~\cite{Blennow:2010th}:
\bee \label{22bbdecay}
m_\text{eff}^{\nu_e}&\simeq&\sum_{i=1}^7 U_{e,i}^2 \,p^2 \frac{m_i}{p^2-m_i^2}\simeq \left(\sum_{i=1}^3 U_{e,i}^2\, m_{\nu_i}\right) \non 
&&+ p^2 \left(- U_{e,4}^2 \,\frac{|m_4|}{p^2-m_4^2}+U_{e,5}^2\, \frac{|m_5|}{p^2-m_5^2}-U_{e,6}^2 \,\frac{|m_6|}{p^2-m_6^2}+U_{e,7}^2 \,\frac{|m_7|}{p^2-m_7^2}\right)\,,
\eee
where $p^2 \simeq - (125 \mbox{ MeV})^2$ is the virtual momentum of the neutrino. 
From the analytical expressions derived in Appendix~\ref{perturbative}, one can see that  
in the limit $\mu_{i,j},\,m_{i,j} \rightarrow 0$,  one has 
$m_5\rightarrow m_4,\, m_7 \rightarrow m_6,\, U_{e,4}^2\rightarrow U_{e,5}^2,\,U_{e,6}^2 \rightarrow 
U_{e,7}^2$,  and thus the extra contribution vanishes.

Our predictions for the effective electron neutrino mass are collected in Fig.~\ref{bbdecay.jpg}, 
and displayed as a function of the mass of the  lightest sterile state, $m_4$. By defining   an "average" effective sterile mass, 
$m_s= \frac{m_4+m_5+m_6+m_7}{4}$,  three distinct mass regimes for $m_s$ can be identified from 
Fig.~\ref{bbdecay.jpg},

\begin{itemize}
\item 
$m_s \ll |p|$: in this regime the effective mass goes to zero, since from Eq.~(\ref{22bbdecay}) one 
approximately has
\be\label{lightsteriles}
m_\text{eff}^{\nu_e}\,=\,p^2 \sum_{i=1}^7 U_{e,i}^2 \, \frac{m_i}{p^2-m_i^2}\,\simeq 
\sum_{i=1}^7 U_{e,i}^2 \,m_{_i}\,,
\ee
and one can write 
\be
\sum_{i=1}^7 U_{\alpha,i}^2 \,m_{_i} \,=\, \sum_{i=1}^7 U_{\alpha,i} \,m_{_i} \,U_{i,\alpha}^T 
\,= \,M_{\alpha, \alpha}\,, 
\ee
where $M$ denotes the full neutrino mass matrix. 
\item $m_s \approx |p|$: the contribution of the pseudo-Dirac states becomes more important, and can induce sizeable effects to $m_\text{eff}^{\nu_e}$.
\item $m_s \gg |p|$: in this regime the heavy states decouple, and the contributions to $m_\text{eff}^{\nu_e}$ only arise from the 3 light neutrino states.
\end{itemize}

Notice that the values of $m_\text{eff}^{\nu_e}$ displayed in Fig.~\ref{bbdecay.jpg} correspond to conservative
(maximal) estimations; since in our scan all parameters are taken to be real, 
no cancellation due to possible (Majorana) phases can take place, and thus reduce the contributions of the "(2,2) ISS" model.
It is important to stress that all points leading to Fig.~\ref{bbdecay.jpg} comply with all available low-energy constraints discussed in Section~\ref{Sec:constrains}. 
The MEG bound on Br($\mu \to e \gamma$)~\cite{Adam:2013mnn} and the constraints from laboratory experiments~\cite{Atre:2009rg} are particularly important, and the latter are in fact responsible for the exclusion of a significant amount of points found (corresponding to the grey regions) in  Fig.~\ref{bbdecay.jpg}.

\begin{figure}[htbp]
 \begin{center}
\includegraphics[width=11.5cm]{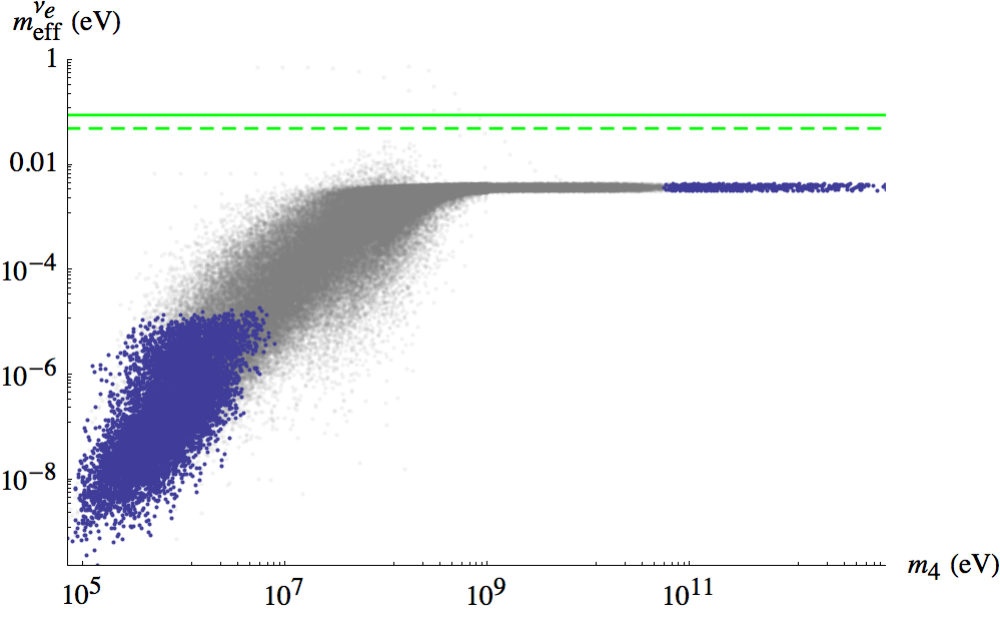}
\caption{Effective electron neutrino mass, $m_\text{eff}^{\nu_e}$, as a function of the lightest sterile mass $m_4$. The green full and dashed horizontal lines denote the current upper bound and the expected future sensitivity \cite{GomezCadenas:2011it}; blue points pass all imposed constraints (oscillation data, NSI, Br($\mu \to e \gamma$) and laboratory direct searches), while grey points are excluded by laboratory bounds. Scan details as in Fig.~\ref{nsi22gev}.
}
\label{bbdecay.jpg}
 \end{center}
\end{figure}

\subsection{The "(2,3) ISS" realisation}
We now address the phenomenology of the next-to-minimal configuration,  the "(2,3) ISS", where two generations of RH neutrinos and three sterile states are added to the SM content. In view of the degree of complexity of the analytical 
expressions derived for the simpler "(2,2) ISS", in this case we directly base our analysis  on a numerical approach.

\subsubsection{Allowed mass hierarchies}
Concerning the neutrino spectra, the crucial difference of the "(2,2) ISS" and the "(2,3) ISS" configurations is that the latter contains {\it four} light states, one being dominantly sterile-like. Its mass typically lies below the GeV (in the analysis we have explored the interval $[0, 100] \ \text{keV}$ for all the entries of the $\mu$ submatrix); recall that the four remaining states are heavy,  pseudo-Dirac pairs.
As can be seen in Table~\ref{massspectrum}, and similar to what occurred for the "(2,2) ISS", the lightest neutrino is also 
massless in the "(2,3) ISS" configurations. Thus, bounds on squared mass differences also translate into bounds for the masses themselves. 

Our study reveals that the "(2,3) ISS" model is not as  fine-tuned 
as the "(2,2) ISS" one. Allowing the entries 
of each submatrix of Eq.~(\ref{generalmatrix}) to vary over one order of magnitude leads to abundant solutions in agreement with low-energy neutrino data. Concerning the hierarchy of the light neutrino spectrum, we have verified that both NH and IH spectra are possible in the explored "(2,3) ISS" parameter space, although IH tends to be only marginally allowed, as is illustrated on 
Fig.~\ref{m1m2sq1.23}. For the left panel (NH), the parameters were varied as
$d_{i,j} \in  [10^{6},10^{7}]$~eV, $n_{i,j} \in  [10^{7},10^{8}]$~eV, $m_{i,j},\,\mu_{i,j} \in  [10^{-1},10]$~eV, while leading to the 
right plot (IH) we considered $d_{i,j} \in  [10^{6},10^{7}]$~eV, $n_{i,j} \in  [10^{8},10^{9}]$~eV, 
$m_{i,j},\,\mu_{i,j} \in  [10,10^3]$~eV.

\begin{figure}[htbp]
 \begin{tabular}{cc}
\includegraphics[width=75mm]{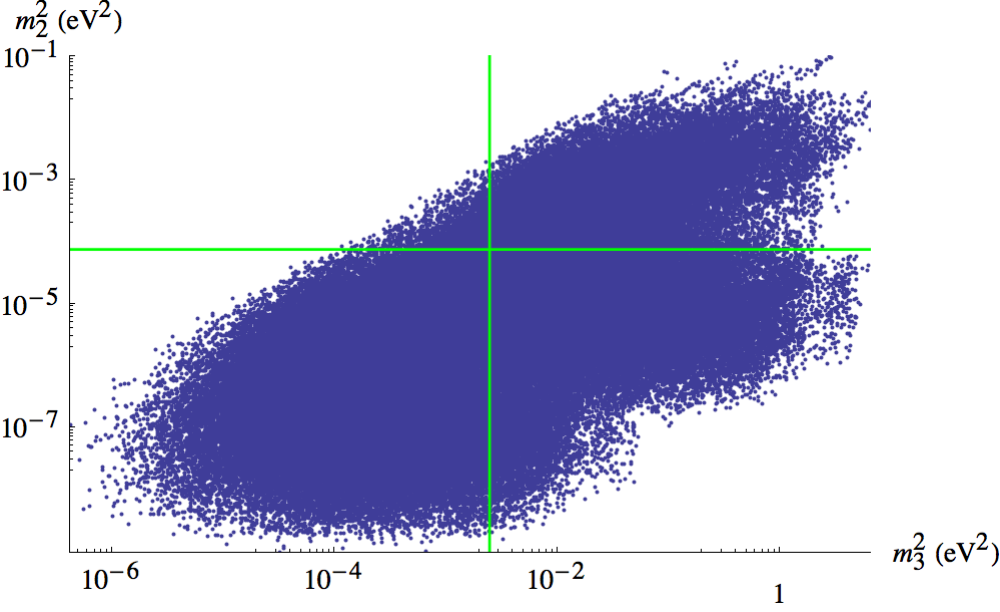}
\hspace*{2mm}&\hspace*{2mm}
\includegraphics[width=75mm]{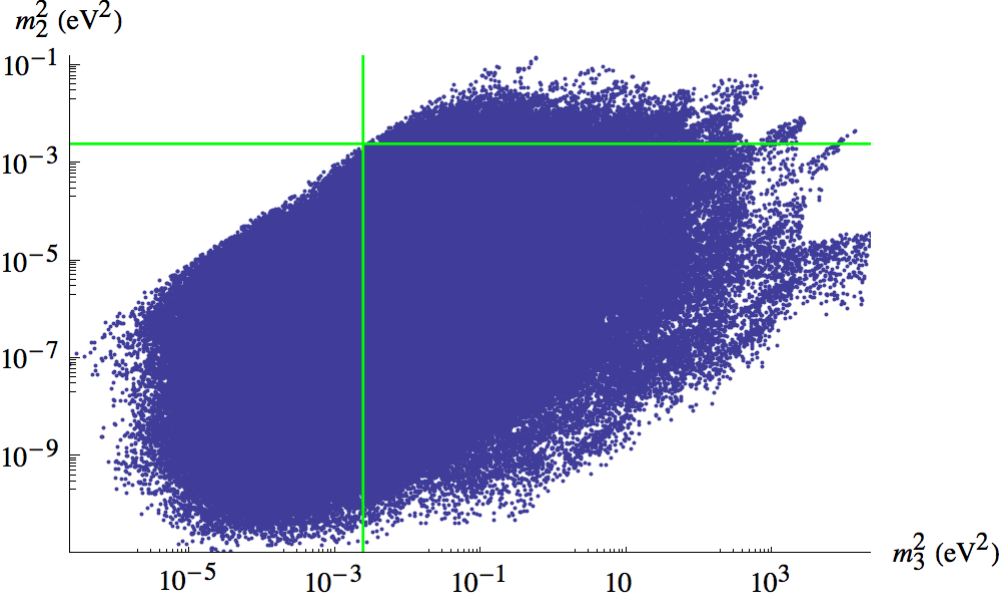}
\end{tabular}
\caption{Squared masses of the active neutrinos for the "(2,3) ISS" model (the lightest neutrino is massless). All points displayed fulfill the experimental constraints on the PMNS entries in the NH (left) and IH (right) schemes. The green lines denote the experimental best fit values \cite{GonzalezGarcia:2012sz} in the NH or IH schemes.
 The scan details are summarised in the text.}
\label{m1m2sq1.23}
\end{figure}

\subsubsection{Constraints from non-unitarity}
Similar to what was previously discussed for the "(2,2) ISS" configuration, the 
constraints coming from the non-observation of NSI (see Section~\ref{Sec:constrains}) also apply to 
"(2,3) ISS" models. We conducted here an analogous study:  the formulae and notations are simple generalizations of those introduced in Section \ref{NSI}, the only difference being that 
in the present case the index $i$ in Eq.~(\ref{nsi}) runs over the states that are integrated out ($\gtrsim 1$ GeV), 
i.e., $i=5,\dots,8$. 
Moreover and since we are interested in a potential "Warm" DM  candidate, we consider realisations of the "(2,3) ISS" model in which only the lightest sterile state lies below $100$ keV (i.e. $\mu\in[0,100]$ keV). 

In Figure~\ref{nsi23gev} we display two examples of deviations from unitarity as parametrised by 
$\epsilon_{\alpha \beta}\equiv \left| \sum_{i=5}^8 U_{\alpha,i} \,U_{i,\beta}^\dagger \right|$ as a function of  an effective  factor $k$. 
We notice that 
the relative density of points in the figure confirms that the "(2,3) ISS" allows for both spectra, although with a clear preference for NH. As in the previous "(2,2) ISS" model, we again verify that 
NSI constraints significantly reduce the number of viable solutions for a "(2,3) ISS" configuration. 
Leading to this figure, we varied  the entries of each submatrix  as $d_{i,j} \in  [10^{3}, 1.7 \times 10^{11}] \text{ eV}$, $n_{i,j} \in  [4.3 \times 10^4,4.8 \times 10^{14}] \text{ eV}$ and  $m_{i,j},\mu_{i,j} \in  [2\times10^{-2}, 10^5] \text{ eV}$. 

\begin{figure}[htbp]
 \begin{tabular}{cc}
\includegraphics[width=75mm]{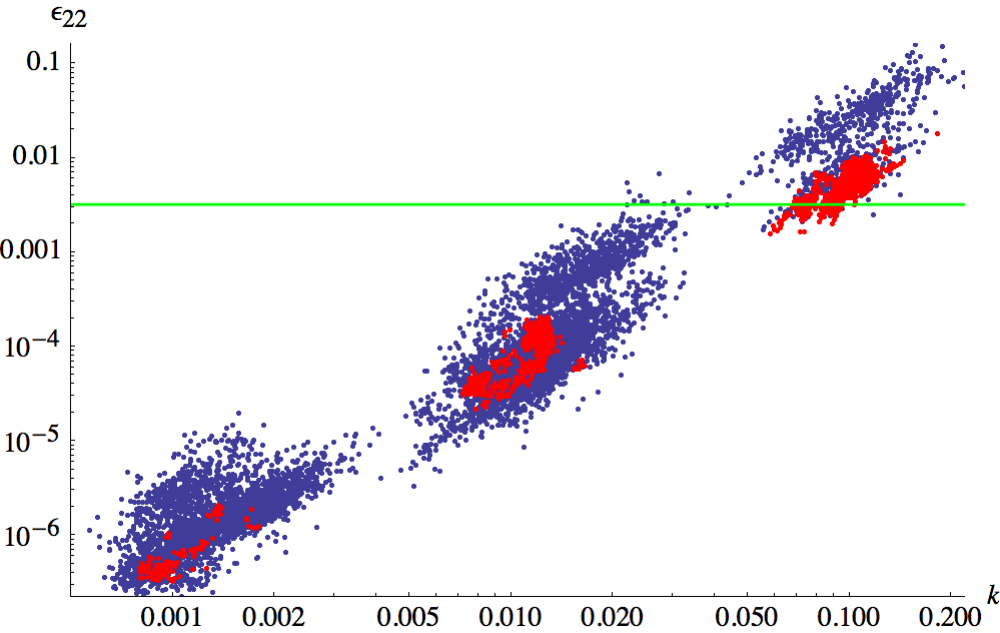}
\hspace*{2mm}&\hspace*{2mm}
\includegraphics[width=75mm]{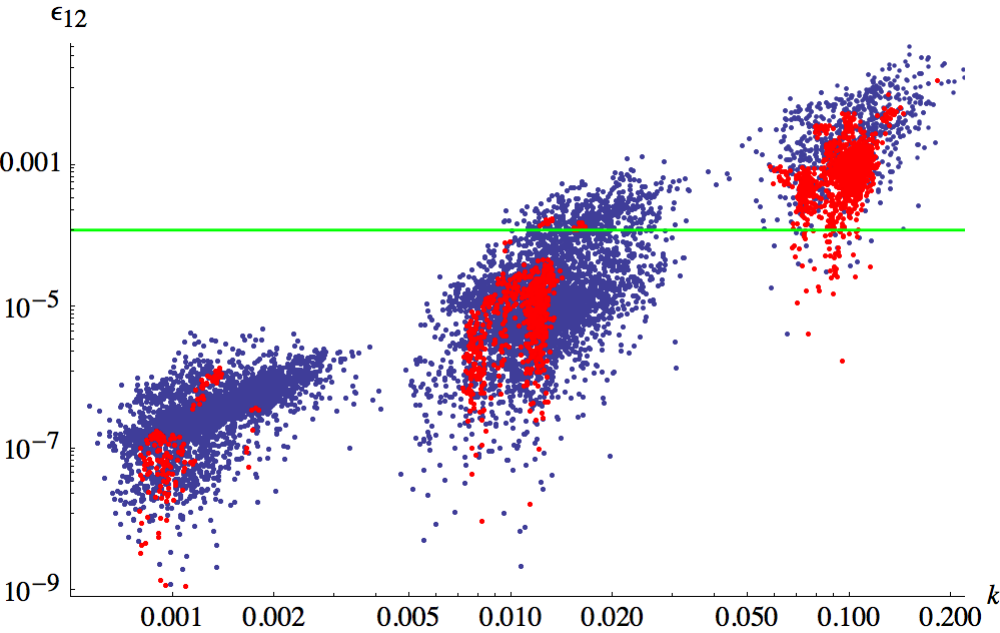}\end{tabular}
\caption{Examples of $\epsilon_{\alpha \beta}\equiv \left| \sum_{i=5}^8 U_{\alpha,i} \,U_{i,\beta}^\dagger \right|$
 entries, as  a function of an effective factor $k$ (generalization of Eq.~(\ref{k}) for the "(2,3) ISS" model).
On the left, $ \epsilon_{22}$, for a  
mass regime in which the sterile neutrino masses are  between 1 GeV and $\Lambda_\text{EW}$; on the right, $ \epsilon_{12}$, in the regime where all sterile states are heavier than  $\Lambda_\text{EW}$.
The green lines indicate the corresponding upper bounds \cite{Antusch:2008fk}. Blue (red) points comply with oscillation data in the NH (IH) scheme. The scan details are summarised in the text.}
\label{nsi23gev}
\end{figure}

\subsubsection{LFV constraints: Br(${\mu \to e \gamma}$)}
For completeness, we illustrate the contributions of the new sterile states to rare LFV processes, in particular    considering Br($\mu \to e \gamma$), see Eq.~(\ref{muegamma}). In Fig.~\ref{fig23muegamma}, we display this observable as a function of the mass of the next-to-lightest sterile state, $m_5$. The investigated parameter space leads to contributions typically below the future experimental sensitivity. However, for $m_5$ in the range $[10^{2}, 10^{4}]$ GeV, one might observe a cLFV signal of the "(2,3) ISS" at MEG.

\begin{figure}[htbp]
 \begin{center}
\includegraphics[width=85mm]{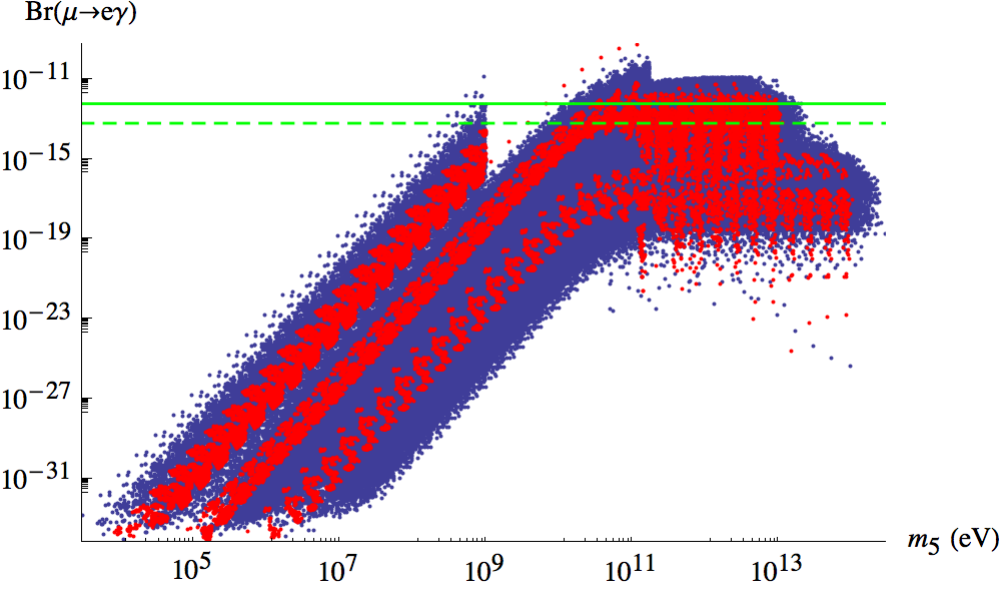}
\caption{Br(${\mu \to e \gamma}$) as a function of the mass of the next-to-lightest sterile state, $m_5$. 
The green full (dashed) horizontal lines denote MEG's current upper bound~\cite{Adam:2013mnn} (future sensitivity~\cite{Baldini:2013ke}); 
blue and red points  correspond to NH and IH solutions, respectively, and 
 pass all imposed constraints (oscillation data and NSI). Scan details as in Fig.~\ref{nsi23gev}.
}
\label{fig23muegamma}
 \end{center}
\end{figure}

\subsubsection{An intermediate sterile scale}\label{lightsterile}
A fundamental difference between the "(2,2)" and the "(2,3) ISS" models is that, since in the latter case $\# s - \# \nu_R = 1$ (see  Section~\ref{generaliss}), the model has a third intermediate energy scale $\mathcal{O}(\mu)$, which corresponds to the mass of a sterile state.  It follows that if $\mu \approx $ eV this model can accommodate a $3\ +\ 1$-scheme that can potentially explain the (anti)-neutrino anomalies in  the short baseline, Gallium and reactor experiments. Should $\mu \approx $ keV, then the model can potentially provide a WDM candidate (see for example the analysis of~\cite{deVega:2013ysa}).

In Figure~\ref{nu_e-sterile_1_23}  we display the mixings of the light sterile state with $\nu_e$, as a function of  $m^2_{4}$. 
All points are in agreement with constraints from oscillation data,  
NSI, laboratory and LFV constraints.
As is clear from Fig.~\ref{nu_e-sterile_1_23}, the parameter space of the "(2,3) ISS" can provide solutions to either reactor anomaly. It can also provide a WDM candidate in the form of a sterile state of mass $\sim 1$ keV. 

\begin{figure}[htbp]
 \begin{center}
\includegraphics[width=10cm]{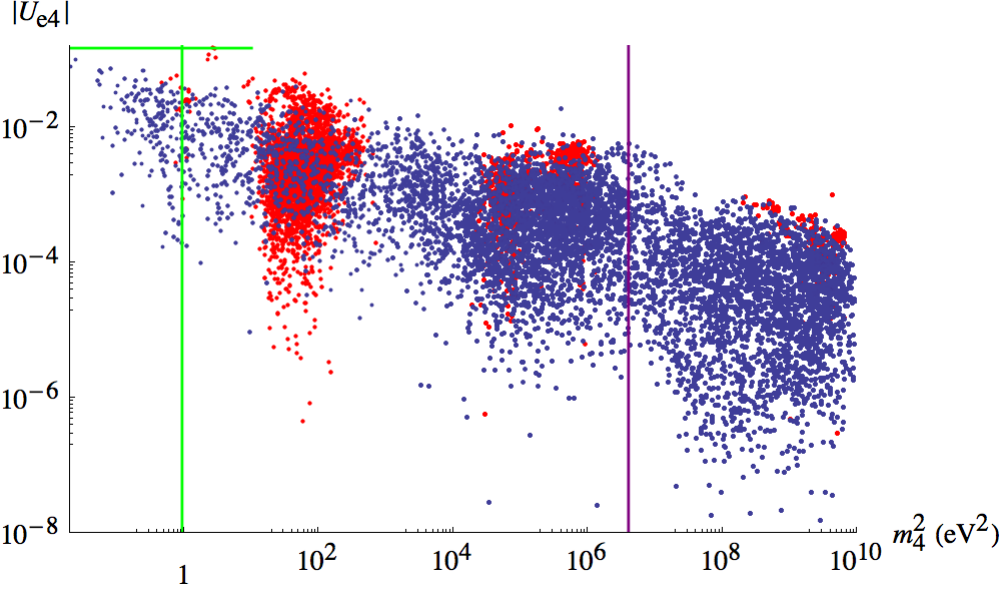}
\caption{Mixings between the electron neutrino and the lightest sterile state, as a function of the sterile squared mass $m_4^2$. The green lines indicate the best fit values of  $(\Delta m_{41}^2,|U_{e4}|)$ for the 3 + 1-scheme~\cite{Kopp:2013vaa}, while the purple vertical  line indicates the value $m_4^2=(2 \text{ keV})^2$, corresponding to the mass of the (warm) dark matter candidate suggested 
in~\cite{deVega:2013ysa}. 
Blue and red  points correspond to NH and IH solutions, respectively. The points displayed comply
with all imposed constraints (oscillation data, laboratory, NSI and Br(${\mu \to e \gamma}$)). Scan details as in Fig.~\ref{nsi23gev}.}
\label{nu_e-sterile_1_23}
 \end{center}
\end{figure}

\subsubsection{Neutrinoless double beta decay}\label{nu0bb-bis}
Due to the presence of the extra light sterile state, in the "(2,3) ISS" model there is an additional contribution to the effective mass derived in Eq.~(\ref{22bbdecay}). In our analysis we assumed the lightest sterile state to have a mass 
$m_4<100 \text{ keV} \ll |p| \approx 125 \text{ MeV}$, it contributes to the neutrinoless double beta  decay effective mass as 
\bee \label{23bbdecay}
m_\text{eff}^{\nu_e}&=&
\sum_{i=1}^8 U_{e,i}^2 \,p^2 \,\frac{m_i}{p^2-m_i^2} \,\simeq \left(\sum_{i=1}^4 U_{e,i}^2\, m_{\nu_i}\right) \non 
&&+ p^2\, \left(- U_{e,5}^2 \,\frac{|m_5|}{p^2-m_5^2}+U_{e,6}^2 \,\frac{|m_6|}{p^2-m_6^2}-U_{e,7}^2 \,\frac{|m_7|}{p^2-m_7^2}+U_{e,8}^2 \,
\frac{|m_8|}{p^2-m_8^2}\right)\,,
\eee 
trivially generalising Eq.~(\ref{22bbdecay}) and where above,  $p^2$ is again the virtual momentum of the propagating neutrino.

In Figure~\ref{bbdecay_23} we summarise our predictions for the effective electron neutrino mass as a function of $m_5$. Like in the previous case, by defining an "average" heavy sterile mass $m_s =\frac{m_5+m_6+m_7+m_8}{4}$,  one can  easily identify the three distinct regimes discussed in Section~\ref{nu0bb} for the "(2,2) ISS" scenario. Especially in regimes of heavier sterile masses (i.e., $m_5 \gtrsim 1 \ \text{GeV}$), the model 
is fairly predictive regarding the $0\nu2\beta$ decays: the value of the effective mass in "(2,3) ISS" scenario lies just below the current experimental bound and within the future sensitivity of ongoing experiments \cite{GomezCadenas:2011it}. 
Somewhat lighter sterile masses could also account for an effective mass within experimental reach, but these solutions are already excluded by the recent MEG bound and by laboratory constraints. 

\begin{figure}[htbp]
 \begin{center}
\includegraphics[width=105mm]{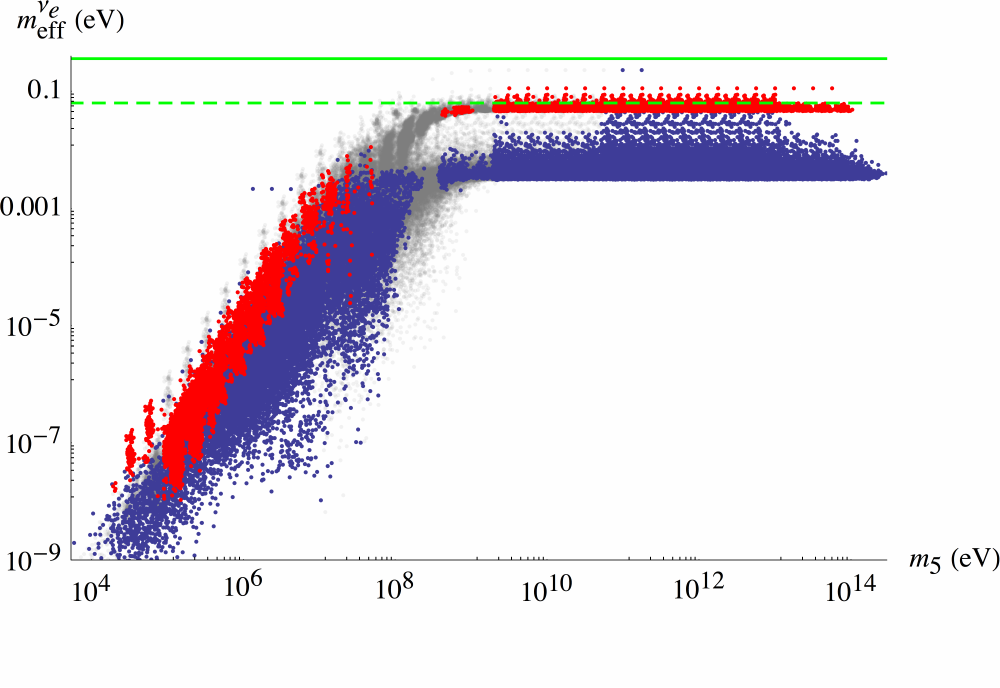}
\caption{Effective electron neutrino mass, $m_\text{eff}^{\nu_e}$, as a function of $m_5$. The green full and dashed horizontal lines denote the current upper bound and the expected future sensitivity \cite{GomezCadenas:2011it}; blue and red  points correspond to NH and IH solutions, respectively, and 
 pass all imposed constraints (oscillation data, NSI, Br($\mu \to e \gamma$) and laboratory direct searches), while grey points are excluded by laboratory bounds. Scan details as in Fig.~\ref{nsi23gev}.
}
\label{bbdecay_23}
 \end{center}
\end{figure}

\section{Conclusions and future prospects}\label{conclusions}

In this work we proposed a methodological approach to identify the most minimal Inverse Seesaw realisations
fulfilling all phenomenological requirements.
By adding extra sterile fermions to the SM (right-handed neutrinos, $\nu_R$, and sterile singlets, $s$)
whose number of generations were not fixed ($\# \nu_R$ not necessarily equal to $\# s$), 
we showed that it is possible to construct several distinct ISS models that can 
reproduce the correct neutrino mass spectrum. 

Our general analysis has shown that the mass spectrum of an ISS model is characterised by either 2 or 3 different mass scales, 
corresponding to the one of the light active neutrinos, that
corresponding to the heavy states, and  an intermediate scale associated to $\#s - \# \nu_R$ sterile states (only relevant 
when $\#s > \# \nu_R$).

The approach we followed was based on  time-independent perturbation theory for linear operators, which allowed to diagonalize the neutrino mass matrix analytically. One can thus obtain analytic expressions for the neutrino eigenstates and associated masses as a power series of the small parameters that violate the total lepton number. 

As a result, we were able to identify two classes of truly minimal ISS realisations that can successfully account for neutrino data.
The first, here denoted "(2,2) ISS" model, corresponds to the SM extended by 
two RH neutrinos and two sterile states. It leads to a 3-flavour mixing scheme, and requires only two scales (the light neutrino masses, $m_\nu$ and the RH neutrino masses, $M_R$). Although considerably fine tuned, this ISS configuration still complies with all phenomenological constraints, and systematically leads to a Normal Hierarchy for the light neutrinos.  
The model could marginally give rise to an effective mass for $0\nu2\beta$ 
within experimental reach, but 
all these regions turn out to be excluded by current laboratory constraints and  MEG bounds on $\mu \to e \gamma$ decays.

The second, the "(2,3) ISS" realisation, corresponds to an extension of the SM by 
two RH neutrinos and three sterile states. This class allows to accommodate both hierarchies for the light spectrum (although the IH  is only marginally allowed), in a  3 + 1-mixing scheme. The mass of the lightest sterile neutrino can vary over a large interval: 
depending on its regime, the 
"(2,3) ISS" realisation can offer an explanation for the 
reactor anomaly (in this case, $m_4 \sim$ eV), or provide a Warm Dark Matter candidate (for a mass of the lightest sterile state around the keV). However, the detailed study of the latter possibility is beyond the scope of this work 
and requires a complete and comprehensive analysis that will be conducted in a subsequent study.   Finally, concerning $0\nu2\beta$ decays, the "(2,3) ISS" scenario leads to effective masses close to the current experimental bound and within future sensitivity of coming experiments~\cite{GomezCadenas:2011it}.

In this work, we have focused on the determination of the truly minimal inverse seesaw  realisations. Our approach can be easily generalised  to probe the phenomenological viability and impact of any ISS extension of the SM (for an arbitrary number of RH states and sterile fermions).

\section*{Acknowledgments}
We are deeply grateful to S.~T.~Petcov  for his valuable and important comments.
We are thankful to J.~P.~Leroy for enlightening discussions. We also thank  C.~Weiland and T.~Schwetz  for their interesting remarks and discussions. We 
 acknowledge  support from the European Union FP7 ITN
INVISIBLES (Marie Curie Actions, PITN-\-GA-\-2011-\-289442).

\setcounter{equation}{0}
\renewcommand{\theequation}{\thesection.\arabic{equation}}
\newpage
\appendix

\section{Perturbative determination of the neutrino masses and of the leptonic mixing matrix}\label{AppendixA}
In the one generation ISS model, and in the basis defined by  $n_L \equiv \left( \nu_L,\nu_R^c,s \right)^T$, 
the neutrino mass matrix can be written as
\be\label{onegenmatrix}
M\,= \,\left( \begin{array}{ccc} 0 & d & 0 \\ d & m & n \\ 0 & n & \mu \end{array} \right),
\ee
where $d,m,n,\mu$ are complex numbers. 
This symmetric matrix can be diagonalized via~\cite{Schechter:1980gr}
\be\label{diagonalization-a}
U^T \,M \,U = \,\text{diag}(m_0,m_1,m_2)\,,
\ee
where $U$ is a unitary matrix and $m_{0,1,2}$ are the physical masses.
To obtain $U$, we use  the hermitian combination $M^\dagger M$ (or $M M^\dagger$),
\be\label{diagmsquare}
\text{diag}(m_0^2,m_1^2,m_2^2) \,= \,\left( U^T \,M \,U \right)^\dagger \,\left( U^T \,M \,U \right)\,  =\, 
U^\dagger \,M^\dagger \,M \,U\,,
\ee
so that the matrix $U$ diagonalizing $M^\dagger \, M$ is the same as the one in Eq.~(\ref{diagonalization-a}).

In the following, we proceed to diagonalize the one-generation  squared mass matrix $M^\dagger\, M$ of Eq.~(\ref{onegenmatrix}), using 
perturbation theory for linear operators. We  also discuss the validity of the perturbative approach. 
The mass matrix $M$ can be decomposed as
\be
M= \underbrace{\left( \begin{array}{ccc} 0 & d & 0 \\ d & 0 & n \\ 0 & n & 0 \end{array} \right)}_{M_0} +\underbrace{ \left( \begin{array}{ccc} 0&0&0\\0&m&0\\0&0&\mu \end{array} \right)}_{\Delta M},
\ee
where $M_0$ is the zeroth order matrix and $\Delta M$ is the perturbation (which violates lepton number by two 
units). One can write $M^\dagger M$ as
\be
M^\dagger M \,=\,\underbrace{M_0^\dagger \,M_0}_{M^2_0} +\underbrace{ \Delta M^\dagger \,M_0 + M_0^\dagger \,\Delta M}_{M_I^2} + \underbrace{\Delta M^\dagger \,\Delta M}_{ M_{II}^2}\,,
\ee
where $M_I^2$ and $M_{II}^2$ are the components of the perturbation that are  homogenous functions of first and second order in the small parameters $ m$ and $\mu$ ($|m|,|\mu| \ll |d|,|n|$).

The perturbativity condition 
$||\Delta M|| \ll ||M_0||$  translates into conditions for the $M_0^2,M_I^2$ and $M_{II}^2$ matrices 
\begin{eqnarray}\label{condnormsI}
\frac{||M_I^2||}{||M_0^2||} &\leq& \frac{2 |m| |d| + 2|m| |n| + 2|\mu| |n|}{|d|^2 + |n|^2}\ll  1 \,,\nonumber \\
\frac{||M_{II}^2||}{||M_I^2||} &\le& \frac{|m|^2 + |\mu|^2}{|m| |n|} \ll 1\,.
\end{eqnarray} 
The perturbative determination of the mass eigenvalues is thus ensuring , 
provided that $|m|,|\mu| \ll |n|$.

For completeness, one must also determine perturbatively
 the matrix $U$ of  Eqs.~(\ref{diagonalization-a}, \ref{diagmsquare}), 
i.e. the leptonic mixing matrix (corresponding to the $U_\text{PMNS}$).
The eigenvalues of $M_0^2$ are given by 
\be
\label{onezeromasses}
\begin{array}{cc}
 {m_0^2}^{(0)} \,=\, 0\,, & {m_{1,2}^2}^{(0)} \,=\, |d|^2+|n|^2\,.
\end{array}
\ee
Denoting by $\mathbf{x}^{(0)}_{0}$ the normalised eigenvector associated to the null eigenvalue and by $\mathbf{x}^{(0)}_1$ and $\mathbf{x}^{(0)}_2$, an orthonormal combination of eigenvectors associated to the degenerate eigenvalue $|d|^2+|n|^2$,  the first order correction to $\mathbf{x}^{(0)}_0$ is given by 
\be\label{eigenvectorcorr}
\mathbf{x}^{(1)}_0 \,=\, \sum_{j=1,2} -\frac{{\mathbf{x}^{(0)}_j}^\dagger M_I^2 \  \mathbf{x}^{(0)}_0}{|d|^2 + |n|^2} \,\mathbf{x}^{(0)}_j\,.
\ee
Since $|\mu|,|m| \ll |n|$,   the coefficients in Eq.~(\ref{eigenvectorcorr}) verify
\begin{eqnarray}\label{condvecpert}
\left| \frac{{\mathbf{x}^{(0)}_j}^\dagger M_I^2\ \mathbf{x}^{(0)}_0}{|d|^2+|n|^2}\right| &\leq & \frac{||\mathbf{x}^{(0)}_j||\ ||M_I^2\ \mathbf{x}^{(0)}_0||}{|d|^2 +|n|^2} \ll 1\,.
\end{eqnarray}

Similar arguments apply to the first order corrections to $\mathbf{x}^{(0)}_{j=1,2}$; the second order 
eigenvector corrections are still subdominant, thus confirming the validity of the perturbative approach.

The lightest neutrino mass arises from perturbative corrections to the $m=0$ eigenvalue, while the two other states are massive and degenerate (pseudo-Dirac heavy neutrinos). The correction to  ${m_0^2}^{(0)}$  at second order is 
\begin{eqnarray}\label{1genneutrinomass}
{m_0^2}^{(2)} &=& \frac{|d|^4 |\mu |^2}{\left(|d|^2+|n|^2\right)^2},
\end{eqnarray}
which reduces to the usual inverse seesaw result once the condition $|d| \ll |n|$ is assumed. 
As discussed in Section~\ref{Sec:towards}, in this approach the only assumption on the magnitude of the physical parameters is driven by the naturalness requirement, i.e.  $|m|,|\mu|\ll |d|,|n|$.

The eigenvector associated to ${m_0^2}^{(2)}$  is given  at zeroth order in the perturbative expansion by\footnote{The phases $\alpha_i$ cannot be fixed by diagonalizing $M^\dagger M$ in (\ref{diagmsquare}). In fact, given an orthonormal basis of vectors, one  can freely change their phases and still have an orthonormal basis. They must be fixed using Eq.~(\ref{diagonalization-a}) and imposing that $m_i \geq 0$ for all $i$.}
\be\label{1genneutrinoeigenvec}
\mathbf{x}_0^{(0)} =e^{i \alpha_0} \left(
\begin{array}{c}
 -\frac{n d^*}{|d| \sqrt{|d|^2+|n|^2}} \\
 0 \\
 \frac{|d|}{\sqrt{|d|^2+|n|^2}}
\end{array}
\right),
\ee
and its  first order correction is
\be
\mathbf{x}_0^{(1)} =e^{i \alpha_0} \left(
\begin{array}{c}
 0 \\
 -\frac{\mu  |d| n^*}{\sqrt{\left(|d|^2+|n|^2\right)^3}} \\
 0
\end{array}
\right). 
\ee
The first order corrections to ${m_{1,2}^2}^{(0)}$ lift the degeneracy of the states and are given by 
\be
\begin{array}{cc}
 {m_1^2}^{(1)} = -\frac{\left|\mu ^* n^2+m |d|^2+m |n|^2\right|}{\sqrt{|d|^2+|n|^2}}, & {m_2^2}^{(1)}= \frac{\left|\mu ^* n^2+m |d|^2+m |n|^2\right|}{\sqrt{|d|^2+|n|^2}},
\end{array}
\ee
with zeroth order eigenstates
\be\label{toyend}
\mathbf{x}_1^{(0)}  =e^{i \alpha_1} \left(
\begin{array}{c}
 -\frac{d^* \left(m |d|^2+m |n|^2+n^2 \mu ^*\right)}{\sqrt{2} \sqrt{|d|^2+|n|^2} \left| {n^*}^2 \mu+m |d|^2+m
   |n|^2\right|} \\
 \frac{1}{\sqrt{2}} \\
 -\frac{n^* \left(m |d|^2+m |n|^2+n^2 \mu ^*\right)}{\sqrt{2} \sqrt{|d|^2+|n|^2} \left| {n^*}^2 \mu+m |d|^2+m
   |n|^2\right|}
\end{array}
\right),
\ee
\be
\mathbf{x}_2^{(0)} =e^{i \alpha_2} \left(
\begin{array}{c}
 \frac{d^* \left(m |d|^2+m |n|^2+n^2 \mu ^*\right)}{\sqrt{2} \sqrt{|d|^2+|n|^2} \left| {n^*}^2 \mu+m |d|^2+m
   |n|^2\right|} \\
 \frac{1}{\sqrt{2}} \\
 \frac{n^* \left(m |d|^2+m |n|^2+n^2 \mu ^*\right)}{\sqrt{2} \sqrt{|d|^2+|n|^2} \left| {n^*}^2 \mu+m |d|^2+m
   |n|^2\right|}
\end{array}
\right).
\ee

\setcounter{equation}{0}

\section{Study of the "(2,2) ISS" realisation}\label{AppendixB}

Here, we use the perturbative approach described above to determine the neutrino spectrum and the leptonic mixing matrix. 
In this minimal model, the neutrino mass terms in the Lagrangian are 
\be
- \mathcal{L}_{m_\nu}\, =\, n_L^T\, C\, {M}\, n_L + \text{h.c.}\,,
\ee
where
\be
\begin{array}{cc}
n_L \,\equiv \, \left(  \nu_L^1,\, \nu_L^2,\,  \nu_L^3,\,  \nu_R^{c,1},\,  \nu_R^{c,2} ,\,  s^1 ,\, s^2 \right)^T,\hspace{1cm} & \text{and}\   C = i \gamma^2 \gamma^0. \end{array}
\ee
The  "(2,2) ISS" mass matrix ${M}$ is given by
\be\label{m22full}
{M} = \left( \begin{array}{ccccccc} 
0&0&0&d_{1,1}&d_{1,2}&0&0\\
0&0&0&d_{2,1}&d_{2,2}&0&0\\
0&0&0&d_{3,1}&d_{3,2}&0&0\\
d_{1,1}&d_{2,1}&d_{3,1}&m_{1,1}&m_{1,2}&n_{1,1}&n_{1,2}\\
d_{1,2}&d_{2,2}&d_{3,2}&m_{1,2}&m_{2,2}&n_{2,1}&n_{2,2}\\
0&0&0&n_{1,1}&n_{2,1}&\mu_{1,1}&\mu_{1,2}\\
0&0&0&n_{1,2}&n_{2,2}&\mu_{1,2}&\mu_{2,2}
 \end{array} \right)\, .
\ee
 Using Eq.~(\ref{nphys}), the number $n_p$ of physical parameters is $24$.  
 In  the following we choose\footnote{The mass matrix of Eq.~(\ref{m22full}) can be cast in such a form through the following  procedure: via a combination of  the transformations  in Eq.~(\ref{sym1.22}) and Eq.~(\ref{sym2.22}), one can always choose a basis in which the charged leptonic mass matrix ${\mathfrak{m}}$ is diagonal and real. With a combined transformation of Eq.~(\ref{sym3.22}) and Eq.~(\ref{sym4.22}) the matrix $n$ can be rendered real and diagonal; similar transformations allow to eliminate two phases form the matrix $\mu$ (for example those in the diagonal) while keeping $n$ real. Finally, another combined transformation of Eq.~(\ref{sym1.22}) and Eq.~(\ref{sym2.22}), allows to make 
one column  of the Dirac mass matrix, $d$, real (the first one, for example), while keeping ${\mathfrak{m}}$ real.} 
a  basis in which one has exactly 24 free parameters, as shown in Table~\ref{physpar22}. 
\begin{table}[htbp]
\begin{center}
\begin{tabular}{|c|c|c|c|}
\hline
Matrix & \# of moduli & \# of phases &  Total \\
\hline
Diagonal and real ${\mathfrak{m}}$ & $3$ & $0$ & $3$ \\
$d$ with one real column & $6$ & $3$ & $9$ \\
 $m$ & $3$ & $3$ & 6 \\
Real and diagonal $n$ & $2$ & $0$ & 2 \\
$\mu$ with real diagonal & $3$ & $1$ & 4\\
\hline
Total & $17$ & $7$ & $24$\\
\hline 
\end{tabular}
\end{center}
\caption{Example of a basis in which the number of parameters matches the number of physical parameters.}
\label{physpar22}
\end{table}

\noindent 
In the chosen basis,  the mass matrices ${M}_0$ and  $ \Delta{M}$ (${M}={M}_0 + \Delta{M}$) are given by
\bea\label{22massmatrix}
\begin{tabular}{cc}
${M} _0= \!\!\left( \begin{array}{ccccccc} 
0&0&0&d_{1,1}&d_{1,2}&0&0\\
0&0&0&d_{2,1}&d_{2,2}&0&0\\
0&0&0&d_{3,1}&d_{3,2}&0&0\\
d_{1,1}&d_{2,1}&d_{3,1}&0&0&n_{1}&0\\
d_{1,2}&d_{2,2}&d_{3,2}&0&0&0&n_{2}\\
0&0&0&n_{1}&0&0&0\\
0&0&0&0&n_{2}&0&0
 \end{array} \right),$&$\!\!\!\!\Delta {M} = \!\!\left( \begin{array}{ccccccc} 
0&0&0&0&0&0&0\\
0&0&0&0&0&0&0\\
0&0&0&0&0&0&0\\
0&0&0&m_{1,1}&m_{1,2}&0&0\\
0&0&0&m_{1,2}&m_{2,2}&0&0\\
0&0&0&0&0&\mu_{1,1}&\mu_{1,2}\\
0&0&0&0&0&\mu_{1,2}&\mu_{2,2}
 \end{array} \right),$
\end{tabular}
\eea
where ($d_{i,1},n_i,\mu_{i,i}$) are real  and ($d_{i,2},\mu_{1,2},m_{i,j}$) are complex numbers.

\subsection{Massless eigenstate}
Having a massless eigenstate is an unavoidable feature of the minimal  "(2,2) ISS" and  "(2,3) ISS" realisations. 
In the minimal  "(2,2) ISS" realisation,  the massless eigenstate is given by 
\hskip -0.3cm\bee
\mathbf{v_1}&=&e^{i \left(\alpha_1-\phi_3\right)} \left( 
\tilde{\Delta_1},
-\tilde{\Delta_2},
\tilde{\Delta_3}, 0\,,0\,,0, 0
\right)^T,\  \tilde{\Delta_i}=\frac{\Delta_i}{\sqrt{|\Delta_1|^2+|\Delta_2|^2+|\Delta_3|^2}} =\left|\tilde{\Delta_i}\right|e^{i \phi_i},\eee
\bee
\text{with} \ \ \Delta_1=d_{2,1}d_{3,2}-d_{2,2}d_{3,1}, \quad
\Delta_2=d_{1,1}d_{3,2}-d_{1,2}d_{3,1}, \quad
\Delta_3=d_{1,1}d_{2,2}-d_{1,2}d_{2,1}, 
\eee
which is compatible with the constraints on the $U_\text{PMNS}$ matrix, in both cases of normal and inverted hierarchy.

\subsection{Perturbative diagonalization}\label{perturbative}
At  zeroth order, the (squared) masses of the system are given by the following set of eigenvalues of the matrix ${M}_0$ of Eq.~(\ref{22massmatrix})
\bee\label{22zeroth}
\lambda  =  \left\{ 0,0,0,\frac{f-\sqrt{f^2-4g}}{2},\frac{f-\sqrt{f^2-4g}}{2},\frac{f+\sqrt{f^2-4g}}{2},\frac{f+\sqrt{f^2-4g}}{2}\right\} ,
\eee

\bee\label{fg}
\text{where}\ \ f&\!\!=&\!\!|d_{1,2}|^2+|d_{2,2}|^2+|d_{3,2}|^2+d_{1,1}^2+d_{2,1}^2+d_{3,1}^2+n_{1,1}^2+n_{2,2}^2,\non
\text{and}\ \ g&\!\!=&\!\! |d_{1,2}|^2 \left(d_{2,1}^2+d_{3,1}^2+n_{1,1}^2\right)+|d_{3,2}|^2
   \left(d_{1,1}^2+d_{2,1}^2+n_{1,1}^2\right)+|d_{2,2}|^2 \left(d_{1,1}^2+d_{3,1}^2+n_{1,1}^2\right)\non 
   &&-d_{1,1} d_{2,1}
   d_{2,2} d_{1,2}^*-d_{1,1} d_{1,2} d_{3,1} d_{3,2}^*-d_{2,1} d_{2,2} d_{3,1} d_{3,2}^*-d_{1,1} d_{3,1} d_{3,2}
   d_{1,2}^*\non 
   && -d_{2,1} (d_{1,1} d_{1,2}+d_{3,1} d_{3,2}) d_{2,2}^*+d_{1,1}^2 n_{2,2}^2+d_{2,1}^2 n_{2,2}^2+d_{3,1}^2
   n_{2,2}^2+n_{1,1}^2 n_{2,2}^2\ .
\eee
Two of the three massless states receive perturbative contributions from 
$\Delta {M}$ of Eq.~(\ref{22massmatrix}) and, at second order in the perturbative expansion, the light neutrino  spectrum is given by
\bee\label{22issmasses}
{m_1^2}^{(2)}=0,\quad
{m_2^2}^{(2)}=\frac{b-\sqrt{b^2+4 c}}{2},\quad
{m_3^2}^{(2)}&=&\frac{b+\sqrt{b^2+4 c}}{2},
\eee
where the parameters $b$ and $c$ are expressed in terms of the entries of the (2,2) mass matrix given in Eq.~(\ref{m22full}) ($b$ and $c$ do not depend on the submatrix $m_{i,j}$). Due to the long and involved expressions for both parameters $b$ and $c$, we refrain from displaying the corresponding formulae here. 
Nevertheless, the compact expressions above allow to extract important information: the "(2,2) ISS" 
scenario strongly prefers the NH scheme.

%%%%%%%%%%%%%%%%%%%%%%%%%%%%%%%%%%%%%

\end{document}